\documentclass[preprint, nolinenumbers, amsmath,amssymb,
 aps, pra, nofootinbib]{revtex4-2}

\usepackage{graphicx}
\usepackage{dcolumn}
\usepackage{bm}
\usepackage{svg}

\def\Neutron{{\text n}}
\def\Hydrogen{\hbox{H}}
\def\Antihydrogen{\hbox{$\overline{\text H}$}}
\def\Muonium{\hbox{Mu}}
\def\Positronium{\hbox{Ps}}
\def\Qn{\hbox{$q_{\text n}$}}

\begin{document}

\title{\textbf{Gravitational and other shifts of whispering-gallery and gravitational state interference patterns of light neutral particles} 
}

\author{V.V. Nesvizhevsky$^1$}
 \email{Corresponding author: nesvizhevsky@ill.eu}
\author{J.A. Pioquinto$^{1,2}$}
\author{K. Schreiner$^{1,3,4,5}$ }
\author{S. Bae\ss{}ler$^{2,6}$ }
\author{P. Crivelli$^{7}$ }
\author{F. Nez$^3$}
\author{S. Reynaud$^3$ }
\author{P. Yzombard$^3$}
\author{S.A. Vasiliev$^8$}
\author{E. Widmann$^{4}$}
 \affiliation{$^1$Nuclear and Particle Physics, Institut Max von Laue - Paul Langevin, Grenoble, France}
\affiliation{$^2$University of Virginia, Charlottesville, Virginia, USA}
\affiliation{$^3$Laboratoire Kastler Brossel, Sorbonne Universite, CNRS, ENS-Universite PSL, College de France, Paris, France}
\affiliation{$^4$Marietta Blau Institut, Austrian Academy of Sciences, Vienna, Austria}
\affiliation{$^5$Vienna Doctoral School in Physics, Vienna, Austria}
\affiliation{$^6$Oak Ridge National Laboratory, Oak Ridge, Tennessee, USA}
\affiliation{$^7$Institute for Particle Physics and Astrophysics, ETH Zurich, Zurich, 8093, Switzerland}
\affiliation{$^8$University of Turku, Turku, Finland}

\date{\today}

\begin{abstract}
We discuss small shifts in the interference patterns of gravitational and whispering gallery quantum states that can be observed with neutrons, atoms, antiatoms, muonium, positronium, and other particles. A gravitational shift of interference patterns of neutron gravitational and whispering-gallery states can be easily observed with cold, very cold, or ultracold neutrons. The developed methods can be used for observing/searching for other shifts in fundamental neutron physics experiments, for instance, for measuring the gravitational constant or constraining the neutron electric charge. A series of such measurements will be made with neutrons at the PF1B/PF2/D17 facilities at the ILL. A peculiar feature of analogous atomic (anti-atomic) experiments is the much smaller effective critical energies of the materials of mirrors for (anti)atoms. We evaluated parameters that make a measurement of the hydrogen and antihydrogen whispering-gallery states and their gravitational shifts feasible. A series of such measurements will be made with hydrogen and deuterium atoms by the GRASIAN collaboration in Vienna and Turku. Such a measurement with antihydrogen atoms may be of interest for the GBAR experiment, the ASACUSA experiment, which is producing a beam of slow antihydrogen atoms, and other experiments at CERN, which study the gravitational properties of antimatter. Quantum reflection of muonium and positronium from material surfaces opens the possibility of observing whispering-galley states, although such measurements remain experimentally challenging. Because of small masses of muonium and positronium, the effective critical energies of the mirror materials are much higher fro them than the effective critical energies for hydrogen and other atoms. The observation of gravitational shifts of such states is particularly demanding because of the extremely short lifetimes of these systems. Measurements of whispering gallery states with all these atoms and particles yield precise and detailed information on the quantum reflection properties of materials, providing valuable input for both fundamental and surface studies.
\end{abstract}

                              
\maketitle

\section{\label{sec:level1}Introduction}

The whispering gallery is a well-known phenomenon observed in a variety of realizations, including sound waves in the air \cite{Strutt,Rayleigh1910,Rayleigh1914,Raman1921} and electromagnetic waves of a broad wavelength range \cite{Mie1908,Debey1909,Budden1962,Liu1997,Oraevsky2002,Vahala2003,Stanwix2005,Grudinin2006,DelHaye2007,Kippenberg2008,Hyun2008,Mendis2010,Chiasera2010,Albert2010}. It consists of wave localization in the vicinity of concave mirrors and is expected for waves of various nature.

The Whispering Gallery States (WGS) of massive particles depend on their masses. A massive particle can become trapped in a potential well, created by the particle's reflection from a curved mirror and its pressing against the mirror's surface because of effective centrifugal force. The interference of quantum states in such a system can lead to the emergence of interference patterns. If these quantum states are long-lived, exploiting such interference patterns can enable very sensitive or precise measurements.
Previously, 
WGS were considered for atoms \cite{Mabuchi1994,Vernooy1997} and antiatoms \cite{Voronin2012} as well as observed for neutrons \cite{Nesvizhevsky2008,Cubitt2009,Nesvizhevsky2010,Nesvizhevsky2011crp,Ichikawa2025} and electrons \cite{Reecht2013}. 

\begin{figure}
    \centering
\includegraphics[width=0.5\linewidth]{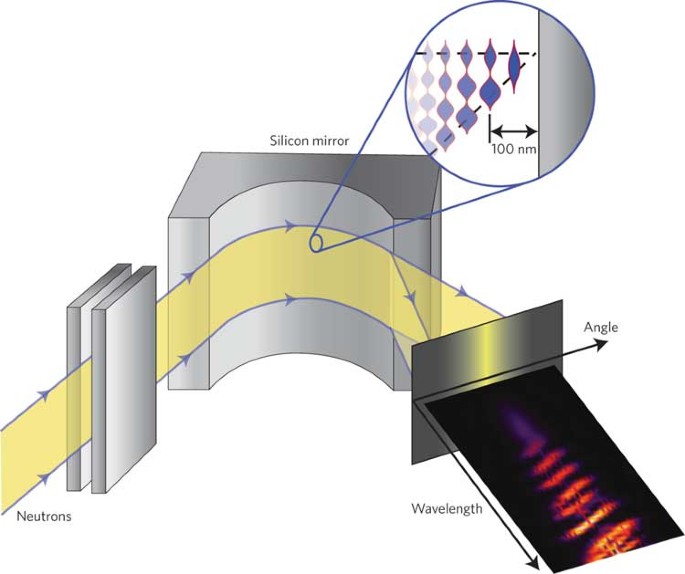}
    \caption{An angularly collimated white neutron beam arrives to the entrance of a concave cylindrical mirror at a small grazing angle to the surface. Some neutrons are trapped in WGS and continue following the surface towards the exit of the mirror. Due to WGS, an interference pattern is observed when measuring simultaneously longitudinal and tangential components of neutron velocities. The longitudinal component of velocity (neutron wavelength) is measured using the time-of-flight technique. The tangential component of velocity (the angle of escape from the mirror edge) is measured in a position-sensitive detector installed at a distance from the exit of the mirror
    (the figure is copied from ref. \cite{Rauch2010}).
    }
    \label{fig:Rauch}
\end{figure}

The WGS method used with neutrons is illustrated in Fig. \ref{fig:Rauch} (copy from ref. \cite{Rauch2010}).

Within a certain simplification, neutrons with an energy of their tangential motion smaller than the critical energy of the mirror material (the neutron-nuclei optical potential \cite{Fermi1936} introduced by Enrico Fermi in 1936) are reflected \footnote{A UCN wavelength is much larger than the interatomic distances in the mirror material, so the interaction is averaged over a huge number of atoms. Due to the electrical neutrality of the neutron, the interaction with electrons is small. Due to the low energy, UCN scattering by a nucleus leads to an isotropic $s$-scattering. Thus, the medium could be thought of as an effective optical potential of $\sim 10^{-7}~eV$, or a nuclear potential of $\sim 10~MeV$ diluted over the volume occupied by the material. The potential can be positive or negative depending on the phase shift of the scattered neutron wave. The shift sign and the potential strength depend on the details of the neutron interaction with each particular isotope.}. This type of neutrons is called ultracold neutrons (UCN). Neutrons with an energy of their tangential motion larger than that are not totally reflected, and subsequently lost. After several quasi-classical bounces, this energy filter conserves subbarrier neutrons and efficiently eliminates the above barrier neutrons.

Due to the approximate equality of the masses of the neutron (\Neutron), hydrogen (\Hydrogen{}) and antihydrogen (\Antihydrogen{}) atoms, the same approach can be applied to the analysis of the feasibility of observing the WGS of \Neutron, \Hydrogen{} or \Antihydrogen{}. In case of \Hydrogen{} and \Antihydrogen{}, a reflecting wall is provided by quantum reflection (QR) on the van der Waals -- Casimir-Polder potential close to the surface \cite{Dufour2013}; note that this potential is attractive, but sharp. QR has been observed and considered for many types of atoms and antiatoms \cite{Lennard1936,Lennard1936bis,Berry1972,Yu1993,Berkhout1993,Carraro1998,Shimizu2001,Druzhinina2003,Pasquini2004,Friedrich2004,Oberst2005,Pasquini2006,Zhao2008}. Although QR of \Antihydrogen{} has not been observed, there is no doubt that it exists because it is based on the same quantum mechanical mechanism as QR of \Hydrogen, and there is no reason to suppose that the electromagnetic interaction is different for \Antihydrogen{} and \Hydrogen{}  \cite{Voronin2005,Froelich2012}. A major difference between neutrons and \Hydrogen{}/\Antihydrogen{} is the relatively small values of the characteristic critical energies of the mirrors for \Hydrogen{} and \Antihydrogen{} compared to those for neutrons. Strictly speaking, the concept of critical energy does not apply to the process of quantum reflection (QR) of atoms and antiatoms. We will define it in the following and use it at the initial stage to explain the choice of parameters for this problem. However, for the actual calculations of the interference patterns, we will use precise calculations of the QR probability for each case studied.

By analogy to the WGS of usual (anti)atoms, two more exotic atoms might also be of interest in this context: muonium (\Muonium, $\mu^+e^-$) and positronium (\Positronium, $e^+e^-$). In the present article, we show that the WGS of \Muonium{} and \Positronium{} should exit and can also be observed. Due to the smaller masses of \Muonium{} and \Positronium{} compared to those of \Antihydrogen{} and \Hydrogen{}, and the correspondingly larger de Broglie wavelengths of these atoms compared to those of \Antihydrogen{} and \Hydrogen{}, a huge increase in the effective critical energy of the mirrors is expected. 

Interference experiments can provide high sensitivity and precision in the cases where this method can be applied. For example, neutrons
in Gravitational Quantum States (GQS) between the gravitational field of the Earth pressing them towards a flat mirror and the reflecting potential of the mirror are useful for searching for exotic interactions and other phenomena beyond the Standard Model \cite{Bertolami2003,Bertolami2005,Blau2006,Baessler2007,Saha2007,Bastos2008,Baessler2009,Nozari2010,Kajari2010,Bastos2011,Pedram2011,Brax2011,Kobakhidze2011,Chang2011,Romera2011,Jenke2011,Brax2013,Abreu2013,Belloni2014,Jenke2014,Pignol2015,Altamirano2018,Anastopoulos2020,Krisnanda2020}. GQS of \Antihydrogen{} 
are of interest in investigating the gravitational interaction of antimatter; the methods could be based on refs. \cite{Nesvizhevsky2002,Jenke2011,Ichikawa2014,Dufour2014,Nesvizhevsky2019,Crepin2019,Tutunnikov2021}.
The characterization of WGS of \Neutron{} and simple atomic systems (SAS) of \Hydrogen, \Antihydrogen, \Muonium{} and \Positronium{} can also be turned to explore these and other phenomena, in particular for searches for new short-range interactions \cite{Dobrescu2006,Antoniadis2011,Safronova2018,Fadeev2019,Kimball2023} or violations of the weak equivalence principle, Lorentz invariance, and CPT symmetry \cite{Kostelecky2011}.
WGS or GQS can also be used to measure the gravitational constant $G$ or to search for the electric charge of the neutron $q_n$.
Finally, the characterization of QR of atoms and antiatoms is a subject of considerable interest in itself. In certain cases discussed in this article, the GQS method cannot be used or is not optimal, and the WGS method is a good alternative.

Small shifts in the WGS interference patterns can be caused by various interactions. For example, a magnetic field gradient will result in a magnetic shift if the particle has a magnetic moment. An electric field will result in an electric shift if the particle has an electric charge. A gravitational field will result in a gravitational shift due to the particle's mass. The weak parity-violating interaction will result in a weak shift. Any interaction of unknown nature can also result in a corresponding shift in the interference pattern, thus allowing one to find or constrain it.

A magnetic shift of the WGS of neutrons has been observed and the sensitivity of such an experiment to a small additional force (of magnetic nature) has been experimentally evaluated \cite{Schreiner2025}. For the minor prototype used and the short measurement time, the ratio of magnetic and centrifugal forces was as small as $\sim 10^{-5}$. The sensitivity of the method might be further increased by many orders of magnitude by properly adjusting the experimental parameters, as explained below. 

The gravitational shifts of the neutron and \Hydrogen{} are relatively easy to measure. A measurement of the gravitational shift of \Antihydrogen{} might be feasible under certain conditions. Due to the very short lifetime of \Muonium{} ($\tau^{Mu}\sim 2.2\cdot 10^{-6}$~s), defined by the lifetime of the muon itself, a gravitational shift of the WGS of \Muonium{} is harder to observe. The lifetime of ortho-\Positronium{} (the triplet state) in the ground state is even shorter ($\tau^{Ps}\sim 1.42\cdot 10^{-7}$~s). However, the lifetimes of highly excited \Positronium{} states (Rydberg states) can be much longer.  They are about twice the corresponding radiative lifetimes of hydrogenic states, owing to the reduced mass of \Positronium{} being half that of the electron--proton system. An important exception is the metastable $2\mathrm{S}$ state, whose decay is still dominated by the $3\gamma$ annihilation channel, with a lifetime scaling as $n^{3}$.  For this state, the total lifetime amounts to $\tau_{2\mathrm{S}}^{\mathrm{Ps}}\sim 1.136~\mu\mathrm{s}$, whereas the purely radiative lifetime would reach about $\sim 2.44\cdot 10^{-1}$~s, i.e. twice the value for \Hydrogen{}.

In Section \ref{sec: parameters}, we present a general method to optimize the WGS parameters of \Neutron, \Hydrogen, \Antihydrogen, \Muonium, and \Positronium.  

In Section \ref{sec: experiment with n}, we propose some applications of the method of precision measurements of neutron WGS shifts, in particular measurements of the gravitational constant $G$ and placing constraints on a neutron electric charge $\Qn$. An observation of a gravitational shift of the WGS of neutrons would provide another example in the "zoo" of quantum phenomena in the gravitational field, in addition to the COW (Colella-Overhauser-Werner) \cite{Colella1975} experiments and gravitational quantum states (GQS) \cite{Nesvizhevsky2002}.

In Section \ref{sec: experiment with H}, we describe a possible scheme for measuring the WGS of \Hydrogen. Note that good conditions are met for the observation of \Hydrogen{} WGS in the experiments of the GRASIAN Collaboration \cite{Killian2023,Killian2024}. Due to the low effective critical energies of any mirror for \Hydrogen, observing the gravitational shift seems feasible. In addition, such experiments with \Hydrogen{} provide almost one-to-one prototyping of possible future experiments with \Antihydrogen.

In Section \ref{sec: experiment with antiH}, we describe a possible scheme for measuring the WGS of \Antihydrogen. Here, the peculiarity of the choice of experimental parameters is to cope with the large velocity of \Antihydrogen, because cooling the \Antihydrogen{} is a rather difficult task.
Such a scheme might be of interest for GBAR, or other \Antihydrogen{} experiments at CERN, in view of the major efforts to directly measure the gravitational interaction of antimatter \cite{Charman2013,Kellerbauer2008,Indelicato2014,Perez2015,Bertsche2018,Anderson2023}. 
The ASACUSA collaboration has recently succeeded in producing an \Antihydrogen{} beam of velocity similar to the existing \Hydrogen\ beams, although with still lower phase space density \cite{hunter:2025}. If the latter can be improved, WGS measurements with \Antihydrogen\ will become possible in a way similar to \Hydrogen.

In Section \ref{sec: muonium}, we describe the WGS of \Muonium. In this case, we optimize the experiment to the parameters of the existing \Muonium{} source \cite{Zhang:2025ogj}.

In Section \ref{sec: positronium}, we describe the WGS of \Positronium. Particular attention is paid to the possibility of observing a gravitational shift of the WGS of \Positronium. 

\section{A method to evaluate parameters of neutron, hydrogen, antihydrogen, muonium, positronium and other WGS}
\label{sec: parameters}

The use of GQS spectrometric and interferometric methods requires sufficiently long observation times of the particles in GQS, which is not always easy to realize in practice because of the high particle velocities or short particle lifetimes.

\subsection{Characteristic parameters of the GQS of neutron, hydrogen, and antihydrogen}

For the particle mass of $\sim 1$ atomic unit (\Neutron, \Hydrogen{} and \Antihydrogen), the characteristic parameters of GQS (the acceleration $a_{GQS}$, the energy of GQS in the ground state $E_{GQS}^{(n,H,\bar{H})}$, the size (height) $l_{GQS}^{(n,H,\bar{H})}$ and the time needed to resolve GQS $\tau_{GQS}^{(n,H,\bar{H})}$ are the following:
  \begin{subequations}
    \begin{align}
a_{GQS}&=g\sim 9.8~\hbox{m/s$^2$} \quad ,
\label{eq: characteristic GQS parameters first}
\\
E_{GQS}^{(n,H,\bar{H})} &=\sqrt[3]{\frac{\hbar^2\cdot m^{(n,H,\bar{H})}\cdot a_{GQS}^2}{2}}\sim 6.0\cdot 10^{-13}~eV \quad ,
\label{eq: characteristic GQS parameters second}
\\
l_{GQS}^{(n,H,\bar{H})} &=\frac{E_{GQS}^{(n,H,\bar{H})}}{m^{(n,H,\bar{H})}\cdot g} \sim 5.9\cdot 10^{-6}~\hbox{m} \quad , 
\label{eq: characteristic GQS parameters third}
\\
\tau_{GQS}^{(n,H,\bar{H})}&= \frac{\hbar}{E_{GQS}^{(n,H,\bar{H})}} \sim 1.1\cdot 10^{-3}~\hbox{s} \quad .
\label{eq: characteristic GQS parameters last}
    \end{align}
  \end{subequations}
We use the standard definition of these characteristic values (\cite{Nes2005}, \cite{Voronin2006}, \cite{Meyerovich2006}, etc). Other definitions could be found in the literature that are different from ours by small factors. $g$ is the gravitational acceleration and $\hbar$ is the reduced Planck constant.   

Let the observation time $t_{GQS}^{(n,H,\bar{H})}$ (the flight time of the particle through the setup, in the flow-through measurement mode) be $\beta_{GQS}^{(n,H,\bar{H})} \sim 10$ times larger than $\tau_{GQS}^{(n,H,\bar{H})}$ (Eq. \ref{eq: characteristic GQS parameters last}), to have the population in the GQS well stabilized:
\begin{equation}
t_{GQS}^{(n,H,\bar{H})}=\beta_{GQS}^{(n,H,\bar{H})}\cdot\tau_{GQS}^{(n,H,\bar{H})}\sim 1.1\cdot10^{-2}~\hbox{s} \quad .
    \label{eq: observation time}
\end{equation}

The difficulty of making and polishing mirrors increases sharply as the size of the mirror increases. For a standard mirror length $L_{GQS}^{(n,H,\bar{H})}$ in neutron GQS experiments \cite{Pignol2007} of:
\begin{equation}
    L_{GQS}^{(n,H,\bar{H})}\sim 0.3~\hbox{m} \quad ,
    \label{eq: mirror length}
\end{equation}
the maximum particle velocity, which meets the condition (\ref{eq: observation time}), is then
\begin{equation}
V_{GQS}^{(n,H,\bar{H})}=\frac{L_{GQS}^{(n,H,\bar{H})}}{t_{GQS}^{(n,H,\bar{H})}}\sim 2.7\cdot 10^1~\hbox{m/s} \quad .
\end{equation}

Increasing the length of the mirror $L_{GQS}^{n,H,\bar{H}}$ by a factor of 2-3 does not change our conclusion: Ultracold particle sources are needed to implement the GQS method.

\subsection{A motivation to use WGS}

Thus, experiments with the GQS are possible with Ultra-Cold Neutrons (UCN) \cite{Lushchikov1969} and with a soft fraction of the spectrum of Very Cold Neutrons (VCN). However, this is an insignificant part of the spectrum of thermal neutron (TN) or cold neutron (CN) sources. The same estimate is valid for the \Hydrogen{} and \Antihydrogen{} atoms. That is, one needs to implement sources of ultra-cold (anti)atoms to increase the fluxes of such particles. 

On the other hand, the WGS method allows for the use of faster particles because the centrifugal acceleration $a_{WGS}$ could be greater than $g$, the energy of WGS, $E_{WGS}^{(n,H,\bar{H})}$, could be greater than $E_{GQS}^{(n,H,\bar{H})}$, thus the characteristic time of formation of WGS, $\tau_{WGS}^{(n,H,\bar{H})}$, might be shorter than $\tau_{GQS}^{(n,H,\bar{H})}$. Therefore, precision spectroscopic and interferometric methods using WGS can be used in a wider velocity range than if using GQS. 

In the following sections, we develop and apply a simple method to optimize WGS and GQS experiments with \Neutron{} and other SAS considered in this work. Both the characteristic parameters of such experiments and the criteria for their optimization differ greatly, so the analysis itself will be performed separately for each case. When discussing the method, we will omit the superscripts corresponding to different particles/atoms but will restore them in the corresponding sections.

\subsection{An optimization method for designing WGS experiments}
\label{sec: optimization method}

For WGS, the effective centrifugal acceleration $a_{WGS}$ is 
\begin{equation}
\label{eq: centrifugal acceleration}
    a_{WGS}=\frac{v_{WGS}^2}{R_{WGS}} \quad ,
\end{equation}
where $v_{WGS}$ is the longitudinal velocity of the particles and $R_{WGS}$ is the mirror radius. By “tuning” the values $v_{WGS}$ and $R_{WGS}$, we can tune the value of $a_{WGS}$. 

Note that by choosing the parameters $v_{WGS}$ and $R_{WGS}$, we can make the centrifugal acceleration greater, equal, or smaller than $g$. However, the magnitude of the acceleration is limited from above by the requirement that the characteristic energy $E_{WGS}$ is well below the effective critical energy of the mirror material, and it is limited from below by the time required to form the corresponding quantum state. The following analysis will be performed separately for different particles.

The characteristic energy $E_{WGS}$ of the WGS is related to the characteristic time $\tau_{WGS}$  of its formation by the uncertainty relation:
\begin{equation}    
\label{eq: uncertainty energy time}
E_{WGS}\sim\frac{\hbar}{\tau_{WGS}} \quad ,
\end{equation}
where $\hbar$ is the reduced Planck constant, and (Eqs. \ref{eq: characteristic GQS parameters second}, \ref{eq: uncertainty energy time})
\begin{equation}
\label{eq: time mass acceleration}
\tau_{WGS}\sim\left(\frac{2\cdot\hbar}{m\cdot a_{WGS}^2}\right)^{\frac{1}{3}} \quad ,
\end{equation}
where $m$ is the mass of the particle. Here we have taken advantage of the fact that the energy differences of low-lying quantum states are similar. 

The characteristic size (height) $l_{WGS}$ of WGS is 
\begin{equation} 
\label{eq: state size}
l_{WGS}\sim\left(\frac{\hbar^2}{2\cdot a_{WGS}\cdot m^2}\right)^\frac{1}{3} \quad .
\end{equation}

In order to observe an interference pattern, one has to populate at least two quantum states. The exact calculation of the spectrum shaping efficiency is a complex task, which depends on the method of shaping the spectrum and the parameters of the problem. In the case of shaping a neutron spectrum in WGS, this problem is analyzed in detail in ref. \cite{Nesvizhevsky2010njf}. For an approximate assessment, the following rule can be used here and in all similar cases below. 
Assume 
\begin{equation}
\label{eq: energy range}
    \gamma_{WGS}\cdot E_{WGS}\sim E_{lim} \quad .
\end{equation}
The ground quantum state has a sufficiently long lifetime if $\gamma_{WGS}\sim 3$. Adding excited states adds $\sim 1$ to $\gamma_{WGS}$ for each excited state (for simplicity, we neglect the decrease in the energy difference between neighbor states with increasing state number).

Note that in the case of the formation of the neutron spectrum in GQS, the spectrum shaping efficiency is analyzed in detail in refs. \cite{Nesvizhevsky2000nima,Nes2005,Voronin2006,Meyerovich2006,Adhikari2007,Nesvizhevsky2010ufn,Escobar2014} (this analysis can easily be generalized to GQS of (anti)atoms since it is not based on the specifics of the interaction of particles with the surface, but only on the geometry of the absorber/scatterer). It can also be easily generalized to the WGS of all particles discussed in this paper. In the following, we will use the same formula (Eq. (\ref{eq: energy range})) and the simple rule for all GQS and WGS.

The QR probability of atoms and antiatoms does not show a sharp threshold behavior. Therefore, the definition of $E_{lim}$ is specific to the problem to be described, in particular the type of (anti-) atoms, the number of quasi-classical bounces and the mirror material. Here, we define it as follows. 

Since in some experiments we measure the anti(atoms) surviving to the mirror edge, and in other experiments we measure the anti(atoms) lost, our criterion will be the condition that the fraction of surviving (anti) atoms is 50\%:
\begin{equation}
\label{eq: critical energy for atoms}
    (P_{QR}(E_{lim}))^{\beta_{WGS}}\sim 0.5 \quad .
\end{equation}

By analogy to GQS, the number of quasiclassical bounces is:
\begin{equation}
\label{eq: number of whispering quasiclassical bounces} 
\beta_{WGS}=t_{WGS}/\tau_{WGS} \quad .
\end{equation}

All that remains to optimize a specific experiment with a specific particle is to select the number of quasi-classical bounces, calculate the effective critical energy for a specific material, and select the energy range. All of these choices can be justified by the goals of particular experiments. All other parameters are related by simple formulas. After such optimization of the experiment, we perform an accurate calculation of the interference pattern and check whether the spatial, temporal, and energy resolutions are really sufficient to perform the experiment.

\subsection{A general method to estimate the statistical sensitivity}

In this paper, we will consider many examples of phenomena and their realizations. The particle fluxes available for them vary by many orders of magnitude. The characteristic parameters of these phenomena also vary by many orders of magnitude. Finally, specific experimental realizations may also vary greatly. Analyzing the statistical accuracy of all these realizations is a labor-intensive task and should be performed in detail elsewhere, case by case. For now, we propose a universal method for assessing the statistical sensitivity of experiments and encourage the reader to perform such an analysis for their own cases. We have conducted such a preliminary analysis and find that all the cases we describe are feasible or might be feasible under certain conditions.

To estimate the number of events that can contribute to the interference pattern, we calculate the corresponding volume of the particle beam's phase space element. The spatial size of the beam is equal to the width of the beam $\Delta Y$ (in the direction parallel to the mirror axis) multiplied by the height of the beam (in the direction perpendicular to the mirror axis). The beam height, in turn, depends on the characteristic size of a quantum state $l_{WGS}$ and the number of quantum states as ($(\gamma_{WGS}+2)\cdot l_{WGS}$) (Eqs. \ref{eq: state size}, \ref{eq: energy range}). The size of the velocity phase space element is usually unlimited in the direction parallel to the beam axis (if the longitudinal velocity is significantly greater than the transverse velocity). In the direction perpendicular to the mirror axis, it is equal to twice the characteristic velocity corresponding to the quantum state energy $E_{WGS}$ multiplied by the square root of the number of quantum states $\sqrt{\gamma_{WGS}}$. The total number of particles in the WGS will be on the order of the fraction of the initial beam flux limited by the phase space element $2\cdot \Delta Y \cdot (\gamma_{WGS}+2) \cdot l_{WGS} \cdot \sqrt{2\cdot E_{WGS}\cdot\gamma_{WGS}/m}$ estimated above, over the total observation time. In order to produce an interference pattern, we need to angularly collimate the incoming beam to be at least a few times, $\delta\sim 5$, smaller than $2\cdot \sqrt{E_{WGS}\cdot\gamma_{WGS}/m}/V_{WGS}$. The total number of counted events must be decreased by the collimation factor $2\cdot \sqrt{E_{WGS}\cdot\gamma_{WGS}/m}/V_{WGS}/\delta$.

\subsection{Modeling of GQS and WGS interference patterns}

To calculate the interference patterns in the following sections, we follow the formalism developed for the calculation of the UCN and \Antihydrogen{} wave functions in the GQS and WGS potentials \cite{Voronin2005, Voronin2006, Voronin2012, Nes2005, Nesvizhevsky2008, Nesvizhevsky2010njf}, the essentials of which we will review in this section. 

The spectrum of WGS and GQS states can be formed in different ways: "too high" energies can be removed by an absorber/scatterer (as is done, for example, in the experiment \cite{Nesvizhevsky2002}) or by the loss of above-barrier quantum states (as is done, for example, in the experiment \cite{Nesvizhevsky2010}); "too low" energies can be removed by a combination of angular collimation of the particle beam and tilting the incident beam relative to the mirror plane (as is done, for example, in the experiment \cite{Nesvizhevsky2010}). For the generality of the theoretical description, we will consider all these cases. Coherence in the incident beam is usually achieved by a sufficiently narrow angular collimation and a sufficiently wide spatial collimation of the beam.

In each experiment, a coherent beam of particles enters a mirror-absorber/scatterer system where the interference pattern is developed. 
For both GQS and WGS experiments, this consists of a mirror of length $L$ for particles to bounce on, a uniform force field generated by the gravitational or centrifugal acceleration experienced by the particle, and a rough absorber/scatterer of length $L_A$ and height $\Delta H$ above the mirror surface. In the proposed experiments, $L_A \leq L$ and the entrance edge of the absorber is placed to coincide with the entrance edge of the mirror. The wave function of a particle propagating along the mirror surface can be expressed as a superposition of quasi-bound states, which interfere with each other. For long-lived particles that reflect well from the mirror surface, the interference pattern can be observed by measuring the flux of particles flowing out of the mirror system with a position-sensitive detector downstream of the mirror. To observe the interference pattern for short-lived particles that decay or annihilate with the mirror surface, it is more feasible to observe the flux of particles interacting with the mirror surface as they bounce. 

To calculate the flux of particles reaching or interacting with the mirror surface, we must calculate the evolution of the wave function $\psi$ of a particle flowing through the setup. To do this, we define the potential for each section of the experiment and solve the Schrödinger equation. For a sufficiently coherent beam, we assume that an incoming particle's wave function can be modeled as a plane wave. In the transverse direction, perpendicular to the mirror surface, the initial wave function is described as 
\begin{equation}
\psi_0 \propto e^{ik_\perp x} \quad .
\end{equation}
Here, $x$ is the altitude of the particle above the mirror and $k_\perp$ is its wave vector perpendicular to the surface. For simplicity, all calculations are performed assuming a zero incidence angle upon the mirror, so $k_\perp = 0$. The longitudinal direction of the wave function is treated classically, so the propagation time experienced by a particle is determined as $t = z/v$ where $z$ is its longitudinal position and $v$ is its velocity. 

The region of the system with a mirror and absorber/scatterer can be described with the potential 
\begin{equation}
V(x) = mgx + V_{mirror}(x) + V_{absorber}(x) \quad .
\end{equation}
For WGS, the gravitational acceleration $g$ should be substituted with the centrifugal acceleration experienced by the particle $a_{WGS}$ from Eq. (\ref{eq: centrifugal acceleration}), with $R_{WGS}$ being the radius of curvature of the mirror.
The exact form of $V_{mirror}$ and $V_{absorber}$ depends on the particle that flows through the system and $V_{absorber}$ on the properties of its (rough) surface. For neutrons with small velocities perpendicular to the surface, small enough such that the neutron optical potential of the mirror is larger than the energy of the neutron's perpendicular motion, the mirror will behave as an infinite potential step. For atoms, polarization and quantum vacuum fluctuations cause interactions between their dipole moments and the mirror, giving rise to the van der Waals-Casimir/Polder potential. Atoms with sufficiently low energy will be repelled by a steep change in potential with high efficiency, a phenomenon known as QR \cite{Dufour2013,Voronin2005}. For low enough energies, this interaction can be well described with the scattering length approximation. In this approximation, the mirror surface potential behaves as an infinite potential step to the incident atoms.
However, it is important to note that the scattering length can have an imaginary part, which physically corresponds to transmission through the van der Waals-Casimir/Polder potential. This plays a key role in the observation of WGS of exotic atoms.

Without the absorber/scatterer, the potential experienced by the particles in each region is a triangular potential well with an infinite potential step at $x = 0$. We can simply model the presence of the absorber/scatterer as another infinite potential step placed at $x = \Delta H$. The rough surface of the absorber/scatterer introduces a new loss mechanism for states propagating through the system which is not accounted for by the step model. We choose to model this effect generically, as was done in \cite{Nes2005, Voronin2006} to effectively describe the observation of GQS of neutrons. When populating states in the presence of the absorber/scatterer, states that have a classical height $x_n = E_n/m g \geq \Delta H$ are assumed to be unpopulated by the end of the absorber/scatterer region, and states with $x_n < \Delta H$ are assumed to pick up imaginary parts $\Gamma_n$ to their energy levels 
\begin{equation}
E_n - i\frac{\Gamma_n}{2} \quad ,
\end{equation}
which determines the lifetime $\tau_n = \hbar/\Gamma_n$ of the state's population. This lifetime can be estimated by 
\begin{equation}
\frac{\Gamma_n}{\hbar} = \omega_n P_n \quad ,
\end{equation}
where $\omega_n$ is the classical bouncing frequency and $P_n$ is the probability of tunneling into the classically forbidden region where the absorber/scatterer resides. For particles that decay, a decaying exponential term multiplies the entire wave function to account for the lifetime of the particle.

In each region of the mirror, the wave function can be expressed as 
\begin{equation}
    \psi(x,t)=\sum_i c_i\psi_i (x,t)e^{-\frac{iE_i t}{\hbar}-\frac{\Gamma_i t}{2\hbar}} \quad ,
    \label{eq: mirror surface wave function}
\end{equation}
where $\psi_i$ are normalized eigenstates of the described potentials and 
\begin{equation}
    c_i = \int\psi(x,t=0)\psi_i(x)dx
    \label{eq: amplitudes}
\end{equation}
are the amplitudes for each state. Note that $\psi_i$ are generically part of a bi-orthogonal set since $\psi_i$ are eigenstates of non-Hermitian Hamiltonians, \cite{Brody2014, Voronin2006}
so $\psi_i$ in Eq. (\ref{eq: amplitudes}) is not conjugated. 
This kind of expression is calculated first for the absorber/scatterer region and then for the absorber/scatterer free region, and the sudden approximation is used to connect the two. 
To calculate the flux of particles flowing out of the mirror system and into a detector far from the system, the probability current $\vec{j}$ of the wave packet entering the free space should be evaluated at the position of the detector. Assuming that the outgoing beam travels in the $\hat{z}$ direction, which corresponds to the tangent line of the surface at the end of the mirror, and the transverse direction of the packet with interference is in the $\hat{x}$ direction, the quantity to calculate is
\begin{equation}
    \vec{j}(x,z)\cdot\hat{z} = \frac{\hbar}{m}\Im{\left(\Psi^*\frac{\partial\Psi}{\partial z} \right)} \quad ,
    \label{eq: detector current1}
\end{equation}
where $z$ is the position of the detector and $x$ is the position on the detector's surface.

In free space, the wave function $\Psi(x,z)$ can be expressed as a superposition of plane waves with the initial wave packet defined by the Fourier transform of the wave packet at the end of the mirror system
\begin{equation}
    \tilde{\psi}(k) = \frac{1}{\sqrt{2\pi}}\int\psi(x,t=T)e^{-ikx}dx \quad ,
\end{equation}
for propagation time $T=L/v$. Then
\begin{equation}
\begin{split}
\Psi(x,z) &= \frac{1}{\sqrt{2\pi}}\int_{-\infty}^{\infty}\tilde{\psi}(k)e^{ikx+i\sqrt{k_0^2-k^2}z}dk \\
&\approx \frac{1}{\sqrt{2\pi}}\int_{-\infty}^{\infty}\tilde{\psi}(k)e^{i\left(-\frac{k^2}{2k_0}z+kx\right)} e^{ik_0z} dk \quad ,
\end{split}
\label{eq: paraxial beam1}
\end{equation}
where $k$ is the wave vector in the $\hat{x}$ direction and $\sqrt{k_0^2-k^2}$ is the wave vector in the $\hat{z}$ direction, and $k_0$ is the magnitude of the total wave vector. In the second line, the paraxial approximation is made since $k<<k_0$. The relevant scale of $k$ is $l_0^{-1}$, so $k^2/2k_0$ has a scale $z_0^{-1}$ where $z_0 = 2k_0 l_0^2$. If $z_0 << z$, then the phase of the exponential in Eq. (\ref{eq: paraxial beam1}) varies rapidly for the relevant $k$ contributing to the wave packet. In the experiments described in the following, $z/z_0 \gtrsim 10^2$ so we can further approximate the wave function by making the stationary phase approximation. The details of the approximation are described in Appendix \ref{sec: Stationary Phase}. The resulting wave function is then 
\begin{equation}
    \Psi(x,z)\approx\sqrt{\frac{k_0}{z}}\tilde{\psi}
\left(k_0 \frac{x}{z}\right)e^{i\left(k_0z+\frac{k_0}{z}\frac{x^2}{2}+\frac{\pi}{4}\right)} \quad .
\label{eq: far field wave packet}
\end{equation}
We must also calculate $\partial \Psi / \partial z$, which is more easily done with Eq. (\ref{eq: paraxial beam1})
\begin{equation}
\begin{split}
    \frac{\partial \Psi}{\partial z}(x,z) \approx& ik_0\frac{1}{\sqrt{2\pi}}\int_{-\infty}^{\infty}\tilde{\psi}(k)\left(1-\frac{k^2}{2k_0^2} \right)e^{i\left(-\frac{k^2}{2k_0}z+kx\right)} e^{ik_0z} dk \\
    \approx& ik_0\frac{1}{\sqrt{2\pi}}\int_{-\infty}^{\infty}\tilde{\psi}(k)e^{i\left(-\frac{k^2}{2k_0}z+kx\right)} e^{ik_0z} dk \quad ,
\end{split}
\label{eq: far field wave packet derivative}
\end{equation}
since $k<<k_0$. Inserting Eqs. (\ref{eq: far field wave packet}) and (\ref{eq: far field wave packet derivative}) into (\ref{eq: detector current1}) produces a current through the detector
\begin{equation}
    \vec{j}(x,z)\cdot\hat{z} \approx \frac{\hbar k_0}{m} \frac{k_0}{z}\left|\tilde{\psi}\left(k_0 \frac{x}{z}\right)\right|^2 \quad .
    \label{eq: detector current2}
\end{equation}
The current in Eq. (\ref{eq: detector current2}) assumes that the beam incident upon the mirror system has a velocity distribution of $n(v)=1$. In reality, the form of $n(v)$ is non-trivial and should be considered in our prediction of the signal. By including $n(v)$ as a multiplicative factor and recognizing that $\hbar k_0 /m=v$, the particle flux $F$ flowing through the detector should be
\begin{equation}
    F(x,v) =n(v) v \frac{k_0}{z} \left|\tilde{\psi}\left(k_0 \frac{x}{z}\right)\right|^2 \quad ,
    \label{eq: free detector flux1}
\end{equation}
and then Eq. (\ref{eq: free detector flux1}) is taken as our signal. For simplicity and generality, $n(v)v=1$ for our calculations. We also drop the explicit $z$ dependence from the argument of $F$ since the detector is typically at a fixed distance. For the most accurate prediction, the interference patterns should be multiplied by the flux spectra $n(v)v$ of the beams of interest. To consider experimental resolution effects, for example, on $x$ and $v$, $F(x,v)$ should be convolved with the appropriate resolution functions $R_x(x)$ and $R_v(v)$ such that the new expected signal is 
\begin{equation}
    \mathcal{F}(x,v) =\int\int F(x',v')R_x(x-x')R_v(v-v')dx'dv' \quad .
    \label{eq: free detector flux2}
\end{equation}

For particles entering a space with a uniform force field, the expected signal is nearly the same as before, but now
\begin{equation}
    F(x,v) =n(v) v \frac{k_0}{z} \left|\tilde{\psi}\left(k_0 \frac{x+\frac{1}{2}at^2}{z}\right)\right|^2 \quad ,
    \label{eq: fall detector flux1}
\end{equation}
where the $x$ coordinate is transformed to follow the trajectories of a classical free fall \cite{Dufour2015} with an acceleration $a$ and free fall time $t = z/v$. Resolution effects can be considered in the same way as before with Eq. (\ref{eq: fall detector flux1}), but now with $F$ taking the definition from Eq. (\ref{eq: fall detector flux1}).

For particles that decay or annihilate on the mirror surface, the signal is taken to be proportional to the probability current of the wave function at the mirror surface $x=0$ and calculated as a function of the longitudinal position $z$ of the particle along the surface. The current entering the mirror surface for the wave function in Eq. (\ref{eq: mirror surface wave function}) is 
\begin{equation}
    \vec{j}(z)\cdot\hat{x} = \frac{\hbar}{m}\Im{\left(\psi^*\frac{\partial\psi}{\partial x} \right)} \Bigg|_{x=0} \quad ,
    \label{eq: mirror current}
\end{equation}
which corresponds to a particle flux
\begin{equation}
    F(z,v) = n(v)\frac{\hbar}{m}\Im{\left(\psi^*\frac{\partial\psi}{\partial x} \right)} \Bigg|_{x=0} \quad .
    \label{eq: mirror flux}
\end{equation}
In calculations of experiments measuring surface current, $n(v)=1$. An exponential decay term is included in Eq. (\ref{eq: mirror flux}) for particles with a lifetime $\tau$. The expected signal then becomes
\begin{equation}
    F(z,v) = e^{-t/\tau }n(v)\frac{\hbar}{m}\Im{\left(\psi^*\frac{\partial\psi}{\partial x} \right)} \Bigg|_{x=0} \quad
    \label{eq: mirror flux decay}
\end{equation}
for $t = z/v$.

Due to our crude treatment of the absorber/scatterer as well as the practical difficulties in understanding its precise behavior in a real experiment, only the probability current after the absorber/scatterer region will be reported. Such complications should not affect strongly the results after the absorber/scatterer region because its primary role is as a state selector, and those states that survive selection are the ones that interact very little with the absorber/scatterer.

\section{Applications of the method of neutron WGS shifts}
\label{sec: experiment with n}

In the experiment to measure the magnetic shift of the WGS of neutrons \cite{Schreiner2025}, the additional acceleration was a factor of $\sim 10^5$ lower than the centripetal and did not differ much from the gravitational one. Therefore, the sensitivity to the gravitational shift could be sufficiently high even with such a mirror with a very large curvature. However, it is obvious that just as the effect itself can increase by many orders of magnitude by increasing the time the neutron stays in the gravitational field, so the sensitivity can be increased by many orders of magnitude by decreasing the magnitude of the centripetal acceleration. We will focus on possible applications of this new method for other experiments in the field of fundamental neutron physics.

The general idea is as follows: an additional force to be measured/detected is applied along the neutron trajectory, that is, before the mirror and/or in the mirror area and/or after the mirror (depending on the conditions of the specific experimental implementation). Such an additional force can, for example, be gravitational, or electrostatic in the case of a search for a hypothetical non-zero electric charge of neutron \Qn. 
One of the strategies to increase the sensitivity is to increase the observation time (trajectory length). The neutron velocity, as well as the mirror radius and the boundary energy of its material, are related parameters and should be optimized together.

\begin{figure}[ht!] 
    \centering
     \includegraphics[width=10cm]{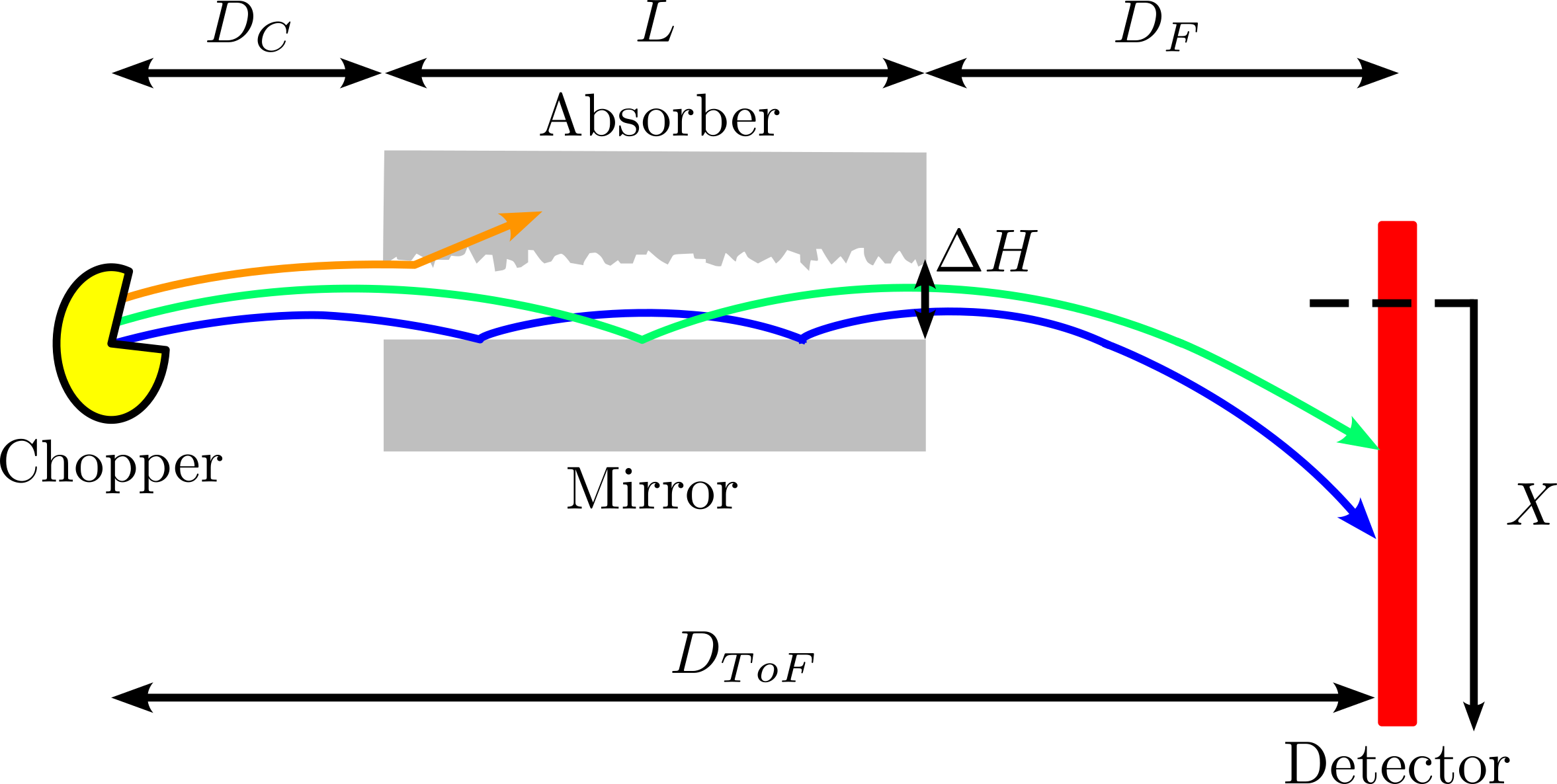}
    \caption{A schematic of a neutron GQS experiment. A collimated beam of neutrons enters the system through a chopper, placed a distance $D_C$ away from the entrance of the mirror, which enables the measurement of the neutrons' velocity by the time of flight (ToF) method. The beam then enters the mirror absorber/scatterer system. Lower energy GQS can pass through the initial section where an absorber/scatterer of length $L$ is placed a height $\Delta H$ above the mirror, while higher GQS are rejected. The lower GQS then propagate along the surface and interfere with each other until reaching the end of the mirror, which has a total length $L$. The exiting beam enters a free fall region where a time and position sensitive detection system is placed a distance $D_F$ from the exit of the mirror. The position of the neutrons on the detector is denoted as $X$.}
    \label{fig: GQS Absorber Interference}
\end{figure}

\subsection{Sensitivity of a GQS interference pattern measured with very cold neutrons to the gravitational acceleration g}
\label{sec: GQS with VCN}

This configuration can be implemented using the equipment that is mostly existing at the ILL, and it is not optimized for any particular purpose. 

\begin{figure}[ht!]
    \centering
     \includegraphics[width=10cm]{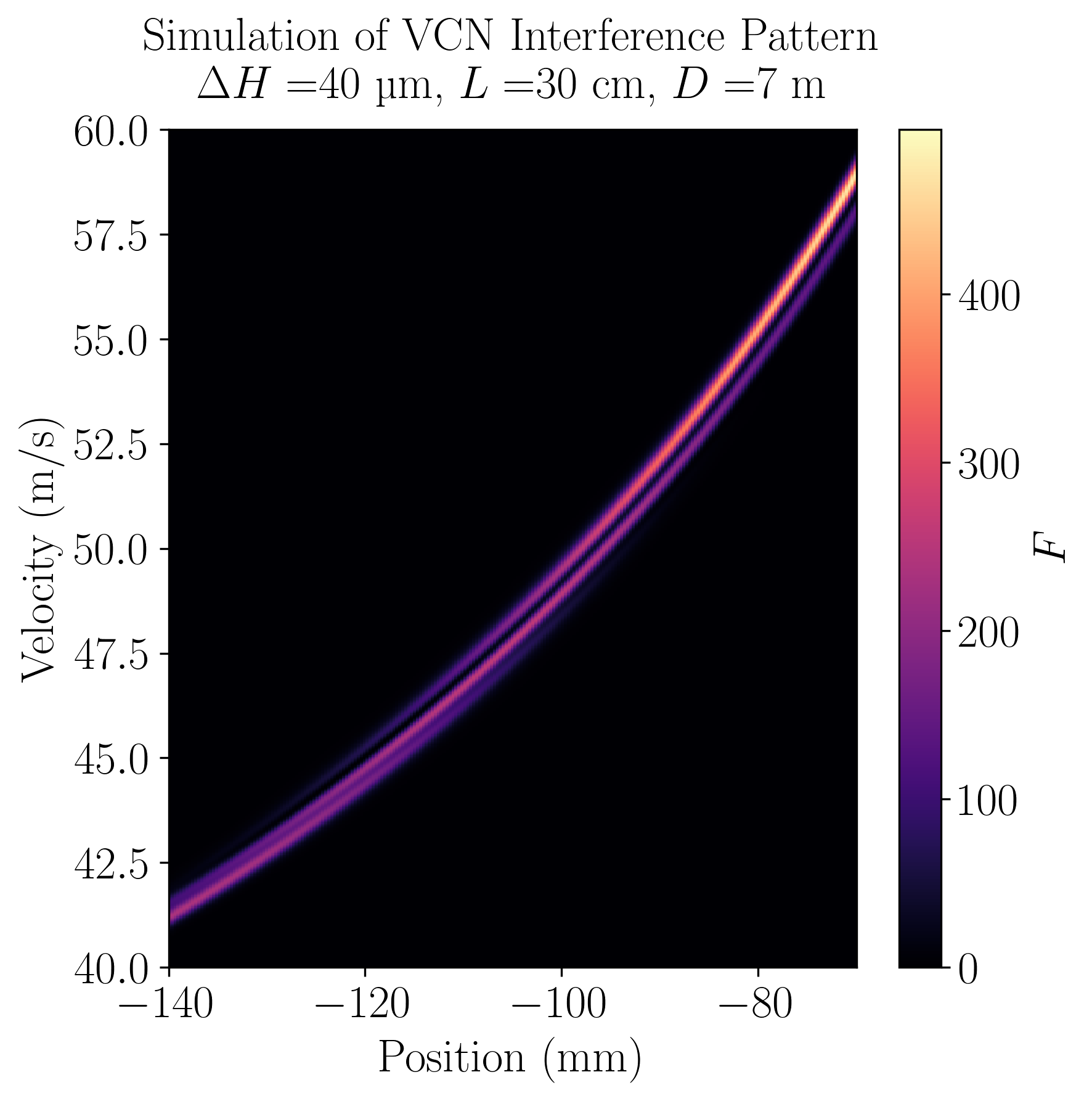}
    \caption{A simulation of the interference pattern generated by propagating VCNs through a mirror-absorber/scatterer system of length $L_{GQS}^{VCN} = 30~\text{cm}$. The height of the absorber/scatterer is set to $\Delta H_{GQS}^{VCN}=40~\mu
    \text{m}$ and the absorber/scatterer is assumed to be very efficient. After exiting the mirror, neutrons enter a free fall region of length $D_{GQS}^{VCN} =7~\text{m}$. The velocities presented in the simulation correspond to the peak in the VCN spectrum at the PF2/VCN instrument. Note that a much richer interference pattern can be produced if a position-sensitive detector with better resolution is available or slower neutrons are used.}
    \label{fig: VCN falling}
\end{figure}

The measurement method is equivalent to that used in ref. \cite{Nesvizhevsky2010}. Very cold neutrons (VCN) come from the PF2/VCN instrument at the ILL; it is proposed to use VCN, not UCN, to increase the statistics and simplify the experiment. Time-of-flight measurement is achieved using a chopper in front of the mirrors. The GQS are formed using a 30~cm long horizontal silicon mirror used in refs. \cite{Killian2023,Killian2024} and a 30~cm long silicon absorber, the operating principle of which is described in ref. \cite{Escobar2014}. The interference pattern can be analyzed using a position-sensitive detector installed at a large distance downstream the mirror, or simply by scanning the neutron intensity with a narrow horizontal slit.
A scheme of this experiment is shown in Fig. \ref{fig: GQS Absorber Interference}, and a simulated interference pattern is shown in Fig. \ref{fig: VCN falling}. We calculated it for the maximum flight path available at PF2/VCN, $D_{ToF} = 7.3$ m. To resolve the interference pattern, the spatial resolution of the position-sensitive detector should not be worse than $\sim 2$~mm. Higher spatial resolution also allows for a shorter flight distance, and thus a reduction in intensity losses associated with the angular divergence of the VCN beam in the horizontal plane. The time window of the chopper should not be greater than $\sim 1.5$~ms. 
The duty cycle of the chopper can be large, up to $\sim 10$\%, because the temporary overlap of the signals in the detector from neutrons of significantly different velocities is effectively eliminated because they are detected at different heights. Note that here and in the following, the insufficient efficiency of the absorber/scatterer can be taken into account (for approximate calculations) using a simple method. Due to the exponentially strong dependence of the probability of losing a particle in a quantum state on the absorber's height, the definition of height can be renormalized: a significantly lower efficiency corresponds to a small decrease in height.

The calculated sensitivity of such an experiment of 10 days to the value of $g$ is $\sim 10^{-4}$. This is the Cramér-Rao bound for this experiment for $N=2\times10^4$ detected neutrons. The description of this calculation can be found in Appendix \ref{sec: Sensitivity} and is calculated with Eq. (\ref{eq: Cramér Rao Bound}), where $P(x,v)=\mathcal{F}(x,v)$ corresponds to the flux simulated using Eqs. (\ref{eq: fall detector flux1}) and (\ref{eq: free detector flux2}). The position resolution function has the shape of a gaussian with two standard deviations corresponding to the stated spatial resolution of the detector. The time resolution is a boxcar function with width equal to the chopper opening time. More realistic resolution functions should be considered on an experiment by experiment basis.

This experiment can be interpreted as a measurement of the weak equivalence principle (WEP), with a neutron being one of the test masses (and the system that is used to determine the local $g$ as the other). This WEP test uses quantum-mechanical properties of the neutron as the test mass; however, its sensitivity does not come close to that obtained in Earth-based ($\sim 10^{-13}$ \cite{Schlamm08}) or satellite-based ($\sim 10^{-15}$ \cite{MICROSCOPE2022}) WEP tests.

Additionally, a measurement of this interference pattern would provide a test for a similar experiment with GQS of \Hydrogen{} that is being carried out by the GRASIAN collaboration \cite{Killian2023,Killian2024}. It would be an intermediate step towards more sensitive experiments described in the following, and it may also provide information on the unexpectedly low GQS population that was observed a few times in previous experiments \cite{Nesvizhevsky2003,Nes2005}.

As the number of GQS (slit height $\Delta H_{GQS}^{VCN}$) increases, the number of interference fringes increases significantly, which means that the sensitivity of the interference pattern increases even more significantly. This improvement is possible with an increase in the spatial resolution of the detector. Note that a thousandfold improvement in spatial resolution has been demonstrated for several types of neutron detectors \cite{Jakubek2009,Jakubek2009bis,Kawasaki2010,Kamiya2020,Clement2022}. However, an even more significant improvement in sensitivity can be obtained using the method presented in the next section.

\subsection{Measurement of the gravitational constant G with an UCN GQS interferometer in reduced gravity}

In order to increase the sensitivity of the method, it seems natural to: a) reduce the longitudinal neutron velocity in order to increase the observation time, b) increase the time neutrons spend in the field, e.g. by providing a longer path length due to multiple specular reflections, and additionally c) reduce the characteristic energy of GQS as the smaller transverse velocity spread allows more specular reflections without overlap of the neutron paths, thus longer observation times.

The decrease in longitudinal velocity is limited by the neutron densities available in the phase space. The maximum UCN velocity is defined by the choice of materials used for the storage and transport systems of the source of UCN and is typically $\sim$6~m/s or less. Thus, the characteristic total UCN velocity is $\sim 4$~m/s, and a component of the velocity is 
\begin{equation}
\label{eq: ucn longitudinal velocity}
V_{GQS^{reduced}}^{UCN}\sim 2~\text{m/s} \quad .
\end{equation}
Even lower UCN velocities would be an advantage in the future. Given the phase space densities of UCN available worldwide, $\sim 10^2$~ UCN/cm$^2$, the fraction of even slower UCN is insufficient for designing interference experiments. However, provided that much higher UCN densities in phase space are available in the future, say $\sim 10^3-10^4$~UCN/cm$^3$, the optimization parameters of such an interferometer may be revised and the sensitivity greatly improved.

\subsubsection{Shaping UCN GQS in reduced gravity}

Reducing gravitational acceleration can be achieved, for example, by tilting the mirror relative to the Earth's gravitational field, as explained in the caption of Fig. \ref{fig:UCN G and qn Schematic}. The reduction is limited by the formation time of the GQS. For a small number of quasi-classical bounces \footnote{We prefer a small number of quasi-classical bounces to access larger effective acceleration values.} 
\begin{equation}
    \beta_{GQS^{reduced}}^{UCN}\sim 3 \quad ,
\end{equation}
and a large mirror length \footnote{This is also to access larger effective acceleration values in the flow-through mode of the pattern shaping.},
\begin{equation}
    L_{GQS^{reduced}}^{UCN}\sim 1~\text{m} \quad ,
\end{equation}
the time of formation of GQS in reduced gravity is
\begin{equation}
\label{eq: formation time in reduced gravity}
   \tau_{GQS^{reduced}}^{UCN}\sim \frac{L_{GQS^{reduced}}^{UCN}}{\beta_{GQS^{reduced}}^{UCN}\cdot V_{GQS^{reduced}}^{UCN}}\sim 1.7\cdot 10^{-1}~\text{s} \quad . 
\end{equation}

This value of $\tau_{GQS^{reduced}}^{UCN}$ (Eq. (\ref{eq: formation time in reduced gravity})) corresponds to the reduced gravitational acceleration (Eqs. \ref{eq: characteristic GQS parameters second}, \ref{eq: characteristic GQS parameters last})   
\begin{equation}
\label{eq: reduced acceleration}
    g_{reduced}^{UCN}\sim 5.1\cdot 10^{-3}~\text{m/s$^2$}
\end{equation}
\footnote{In principle, a much smaller effective acceleration and greater sensitivity of the interferometer can be achieved by using the accumulative mode to shape the initial quantum state, where the experimental challenge consists of the precise installation and control of the mirrors \cite{Pignol2007}. This issue should be better discussed in detail later, provided that the flow-through method for shaping the quantum state is demonstrated and experimentally studied. } and the angle of inclination of 
\begin{equation}
\label{eq: inclination angle}
    \Delta\Theta\sim 0.52~mrad \quad ;
\end{equation}
the angle is the inclination of a normal to the mirror relative to the horizontal plane defined by gravity. 

The characteristic energy of such a quantum state (Eq. (\ref{eq: characteristic GQS parameters second})) is 
\begin{equation}
    E_{GQS^{reduced}}^{UCN}\sim 3.9\cdot 10^{-15}~\text{eV} \quad .
\end{equation}

The corresponding (tangential) velocity is
\begin{equation}
\label{eq: ucn transversal velocity}
    (V_{GQS^{reduced}}^{UCN})_\perp\sim 8.6\cdot 10^{-4}~\text{m/s} \quad .
\end{equation}

The characteristic size of the GQS in reduced gravity is (Eqs. (\ref{eq: characteristic GQS parameters third}, \ref{eq: reduced acceleration}))
\begin{equation}
\label{eq: reduced state size}
    l_{GQS^{reduced}}^{UCN}\sim 7.2\cdot 10^{-5}~\text{m} \quad .
\end{equation}

The exact calculation of the required slit height between the mirror and the absorber/scatterer, as well as the exact calculation of the absorber/scatterer efficiency, is a complex task, analyzed in detail, for example, in refs. \cite{Nes2005,Voronin2006,Meyerovich2006,Adhikari2007,Westphal2007,Nesvizhevsky2010ufn,Escobar2014}.
In these same refs., one can find conditions for choosing the parameters of the absorber/scatterer (in particular, the amplitudes of the surface roughness). 
For an approximate assessment of this value, one can use the same rule as in Eq. (\ref{eq: energy range}) and the paragraph below this equation. The size of the slit corresponding to the passage of one quantum state is $\sim 3l_{GQS^{reduced}}^{UCN}$. Adding each next quantum state adds $\sim l_{GQS^{reduced}}^{UCN}$ (Eq. (\ref{eq: reduced state size})).

\subsubsection{Storing UCN on specular trajectories}

The design of the proposed experiment is inspired by ref. \cite{Borisov1988} in which the storage of UCN on specular trajectories was used to search for the electric charge of the neutron (\Qn). If \Qn$\neq$0, the neutron trajectory would slightly deviate in a strong electric field.  

\begin{figure}[ht!]
    \centering
     \includegraphics[width=11cm]{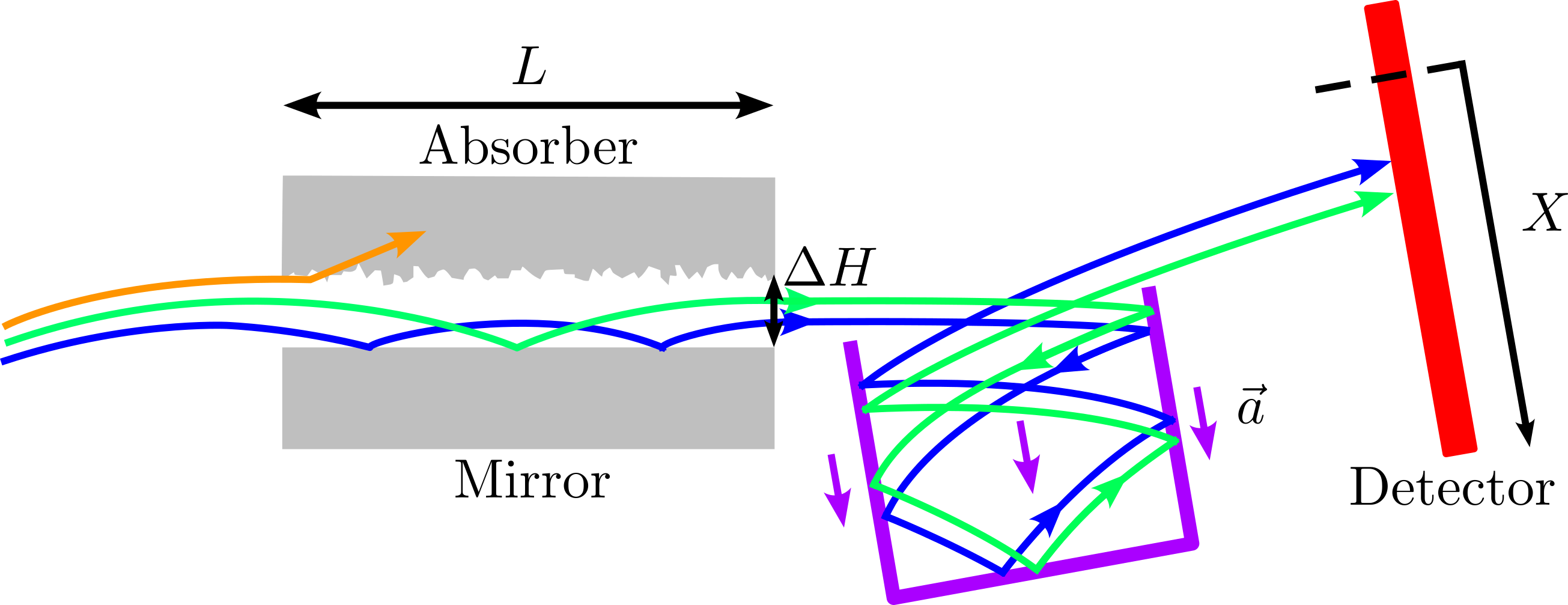}
    \caption{A schematic of the UCN GQS experiment in reduced gravity to measure \Qn~or $G$. An incoming UCN beam enters a mirror absorber/scatterer system which is oriented such that the Earth's gravitational field is pointed nearly normal to the plane of the page. The value of the remaining gravity is defined by the projection of the gravity vector to the plane of the page. GQS with reduced gravity are populated and generate interference patterns which then enter a storage volume where an additional force is exerted on the outgoing beam, either gravitational from a large mass or electric from electrodes, both are generically indicated by $\vec{a}$. After some time in the volume, the beam leaves and reaches a detector to record the interference pattern. An element to shape and analyze the neutron's velocity may also be advantageous, depending on the available UCN spectrum. The optical system for shaping the GQS in reduced gravity can consist of one or more systems of the mirror and the absorber/scatterer (if the characteristic size of the interference structures on the detector is larger than the size of one slit, or, alternatively, if the slits are oriented in a way to direct the beams to the same spot in the position-sensitive detector).}
    \label{fig:UCN G and qn Schematic}
\end{figure}

The optical system for storing UCN on specular trajectories may be a rectangular box with specular internal walls and slits for the input and output of the UCN beam as shown in Fig.
\ref{fig:UCN G and qn Schematic} \footnote{Using a more complex mirror geometry could allow for longer UCN storage on specular trajectories, and consequently, longer UCN trajectories. However, we believe the correct strategy for this experiment is to gradually increase the storage time and, consequently, the sensitivity, so discussing more complex configurations is not currently relevant.}. The storage time of UCN on specular trajectories is limited by the probability of off-specular reflections from the mirror walls. It was studied experimentally and did not exceed $\sim 2\cdot 10^{-3}$ for polished surfaces \cite{Nesvizhevsky2006nima} and $\sim 5\cdot 10^{-3}$ for diamond-like coatings \cite{Nesvizhevsky2007nima}. For polished mirrors and a conservatively estimated frequency of wall collisions
\begin{equation}
    \nu_{GQS^{reduced}}^{UCN}\sim 5 \text{ s}^{-1} \quad
\end{equation}
(which depends on the precise experimental design and the UCN spectrum), the total observation time is limited to
\begin{equation}
\label{eq: time limit for neutron charge}
    t_{GQS^{reduced}}^{UCN}<100~\text{s} \quad.
\end{equation}

However, for the geometry shown in Fig. \ref{fig:UCN G and qn Schematic}, the observation time is also limited by the size of the interference pattern. During this interval (Eq. (\ref{eq: time limit for neutron charge})), the size of the interference pattern will increase too much. Therefore, we will limit the observation time to such a value at which the interference patterns do not yet overlap during successive reflections from mirrors oriented (almost) perpendicular to the UCN motion trajectories. For the widths of these mirrors equal to $1$~m, the longitudinal UCN velocities of $2$~\text{m/s} (Eq. (\ref{eq: ucn longitudinal velocity})) and the transverse UCN velocities of $8.6\cdot 10^{-4}$~\text{m/s} (Eq. (\ref{eq: ucn transversal velocity})), the observation time is equal to
\begin{equation}
\label{eq: time for neutron charge}
    t_{GQS^{reduced}}^{UCN}\sim 20~\text{s} \quad.
\end{equation}
the trajectory length is
\begin{equation}
    D_{GQS^{reduced}}^{UCN}=40~m \quad ,
\end{equation}
and the interference pattern size is  
\begin{equation}
    {\Delta x_{\perp}}=2V_{GQS^{reduced}}^{UCN}\cdot T_{GQS^{reduced}}^{UCN}\sim 3.5~\text{cm} \quad.
\end{equation}

\begin{figure}[ht!]
    \centering
     \includegraphics[width=14cm] {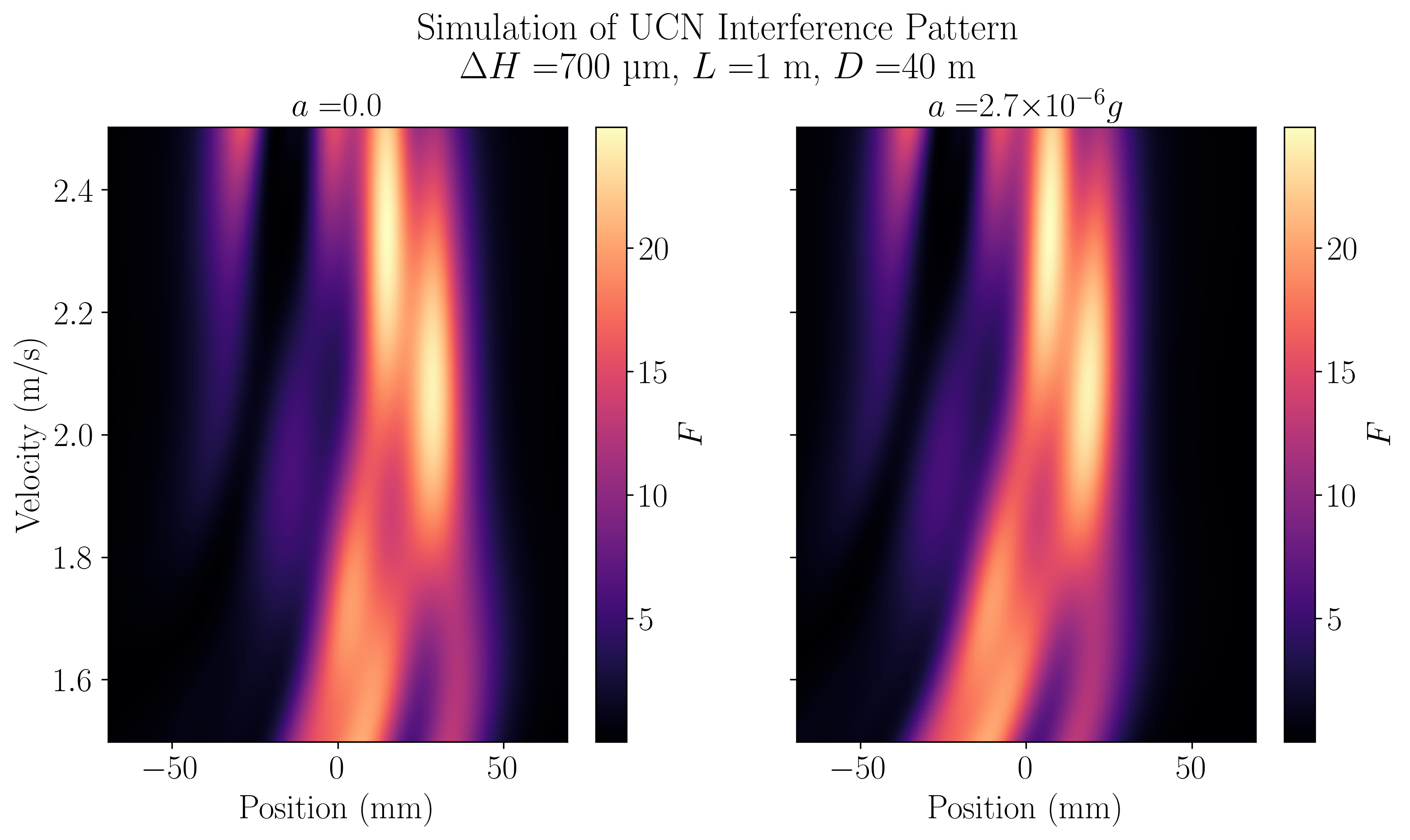}
    \caption{\label{fig:UCN_bigG} Simulations of the expected interference pattern for UCN $G$ and \Qn{} measurements. A mirror absorber/scatterer system of length $L_{GQS^{reduced}}^{UCN}=1$~m is used for calculations with the absorber/scatterer height of $\Delta H_{GQS^{reduced}}^{UCN}=700$~$\mu$m. The system oriented so that the slit between the mirror and absorber/scatterer is nearly vertical, as described in the text. The exiting wave packet is propagated for $D_{GQS^{reduced}}^{UCN}=40$~m, corresponding to the $T_{GQS^{reduced}}^{UCN}=20$~s observation time for $V_{GQS^{reduced}}^{UCN}=2$~m/s UCNs. The effect of a gravitational field from a test mass, or the electric field of electrodes, was included with the presence of a uniform field that allows the exiting wave packet to "fall." The left plot shows the pattern without the effect of an external force and the right has an applied acceleration of $a = 2.7\times10^{-6} g$. This is 10 times higher than the proposed additional acceleration to improve the visibility the pattern's shift to more negative positions.  
     }
    \label{fig: G and qn}
\end{figure}

The simulated interference patterns are shown in Fig. \ref{fig: G and qn}.

\subsubsection{Estimation of sensitivity of the UCN WGS interferometer to the gravitational constant $G$}

The following sensitivity estimate is made under the most general assumptions and serves only as an initial guide. The result depends on the degree of monochromaticity of the longitudinal velocities of UCN, but not very strongly, and the window of longitudinal velocities can be quite wide.
 Suppose that our measurement is the difference in the interference pattern with or without an additional source mass that interacts with the neutrons only by gravitation. For the source mass being a tungsten ball with a diameter of $\sim 1$~m and the UCN just close to the beam, we expect an additional acceleration of $\sim 2.7\cdot 10^{-7}$~g.

Simulations show that with the parameters presented above, $\sim 10$ days of statistics taking, and the phase space density of UCN available at the PF2 facility at the ILL, the sensitivity can reach a few times 10$^{-3}$ of the value $G$. 
In view of the rather contradictory results of existing experiments \cite{CODATA2022}, an independent reliable measurement with a sensitivity better than 10$^{-4}$ may be of interest.
Due to the "fine tuning" of the experimental parameters (mirror sizes, UCN storage time on specular trajectories, the number of quantum states in the interference pattern formation system, statistics collection time, use of future sources with higher UCN density, etc.), such an increase in sensitivity seems realistic. The use of a quantum system can limit some systematic errors. However, the analysis of such an experiment is a separate and extremely complex task, far beyond the scope of this work.

\subsection{Estimation of sensitivity of the UCN WGS interferometer to the neutron electric charge \Qn}

The same comment should be made for the case of assessing the sensitivity of this method to the presence of an electric charge on the neutron \Qn. In order to estimate the sensitivity of the experiment to the presence of \Qn, we assume that the electric field is applied along the entire trajectory of the neutron and its intensity is $\sim 10^4$~V/cm. On the one hand, the electrodes cannot be placed near the mirrors (otherwise, the shadow effect will occur, as the electrodes would be placed in the UCN beam).
However, the field intensity along that part of the neutron trajectory where the interference pattern is small can be many times higher. Therefore, our sensitivity estimate is very conservative and is $\Qn\sim 10^{-22}~e$.
At the same time, all the arguments about the possible "fine-tuning" of the parameters and the further increase in sensitivity by at least one or two orders of magnitude remain valid. In view of the fact that the best limit on the \Qn{} value today is $\sim 10^{-21}~e$ \cite{Baumann1988}, our estimate seems very promising. 

If one decides to take more concrete steps in these directions, the exact experimental design must be developed, the sensitivity optimized, and systematic effects studied theoretically and experimentally. \footnote{An interferometer of this type can also be designed for ultracold atoms, however, the analysis of this possibility is far beyond the scope of this article.} 

\section{A possible measurement of the WGS of Hydrogen and its gravitational shift}
\label{sec: experiment with H}

\subsection{Optimization of experiments with the WGS of \Hydrogen{}}

The effective critical energies of all materials for \Hydrogen{} are quite small, and even smaller for heavier atoms  \cite{Dufour2013,Dufour2013nano,Crepin2017,Crepin2019,Crepin2019hi}. Therefore, we are interested in a material with a maximum effective critical energy. Among those considered in ref. \cite{Dufour2013}, we select Silica. We ignore the liquid-helium surface on the curved mirror, due to experimental difficulties. 

The density in the phase space of the \Hydrogen{} beam can be quite high compared to that of the neutron beams. However, the angular spread in the interference pattern is small simply because of these low effective critical energies. Therefore, we prefer to use quite a lot of quasi-classical bounces and only a few WGS. This choice will provide a sufficiently informative interference pattern with a large number of interference fringes. For this choice, the requirements for the spatial resolution of the \Hydrogen{} detector are less demanding.
 
A motivation for the informative interference pattern is the possibility of extracting unprecedentedly precise and detailed information about the QR process. Thus, for
\begin{equation}
    \label{eq: hydrogen number of bounces}
    \beta_{WGS}^H\sim 10
\end{equation}
and
\begin{equation}
    \label{eq: hydrogen number of states}
    \gamma_{WGS}^H\sim 3+N
\end{equation}
($N$ is the number of exited WGS),
\begin{equation}
    E_{lim}^H \sim 1.2\cdot 10^{-11}~\text{eV} \quad .
    \label{eq: hydrogen critical energy}
\end{equation}
Here, we applied formula (\ref{eq: critical energy for atoms}) to the data calculated in ref. \cite{Dufour2013}. The precise simulations in the following will illustrate the sensitivity to these parameters.

Eqs. (\ref{eq: characteristic GQS parameters first}) to (\ref{eq: characteristic GQS parameters last}), (\ref{eq: energy range}) and (\ref{eq: hydrogen number of bounces}) to (\ref{eq: hydrogen critical energy}) result in 
\begin{equation}
\tau_{WGS}^H=\tau_{GQS}^H\frac{\gamma_{WGS}^H\cdot E_{GQS}^H}{E_{lim}^H}\sim 2.5\cdot 10^{-4}~\text{s} \quad .
    \label{eq: hydrogen time WGS}
\end{equation}

Eqs. (\ref{eq: time mass acceleration}) and (\ref{eq: hydrogen time WGS}) result in 
\begin{equation}
    a_{WGS}^H\sim g\cdot \left(\frac{\tau_{GQS}^H}{\tau_{WGS}^H}\right)^{1.5}\sim 9.0\cdot 10^1~\text{m/s$^2$} \quad .
    \label{eq: hydrogen WGS acceleration}
\end{equation}

To have the mirror compatible with the GRASIAN setup \cite{Killian2023,Killian2024} we select the same mirror length as that in the GQS experiments:
\begin{equation}
    \label{eq: hydrogen WGS mirror length}
    L_{WGS}^H=0.3~\text{m} \quad .
\end{equation}

Eqs. (\ref{eq: hydrogen number of states}), (\ref{eq: hydrogen time WGS}) and (\ref{eq: hydrogen WGS mirror length}) result in 
\begin{equation}    V_{WGS}^H\sim\frac{L_{WGS}^H}{\beta_{WGS}^H\cdot\tau_{WGS}^H}\sim 1.2\cdot 10^2~\text{m/s} \quad .
    \label{eq: hydrogen WGS velocity}
\end{equation}

Eqs. (\ref{eq: hydrogen WGS acceleration}) and (\ref{eq: hydrogen WGS velocity}) result in
\begin{equation}    R_{WGS}^H\sim\frac{(V_{WGS}^H)^2}{a_{WGS}^H}\sim 1.6\cdot 10^2~\text{m} \quad .
    \label{eq: hydrogen WGS radius}
\end{equation}

The ratio of the magnitudes of gravitational and centrifugal accelerations (Eq. (\ref{eq: hydrogen WGS acceleration}))
\begin{equation}
    \frac{g}{a_{WGS}^H}\sim 1.1\cdot 10^{-1}
    \label{eq: hydrogen gravitational shift}
\end{equation}
is “huge” compared to the $\sim10^{-5}$ value measured in the neutron experiment \cite{Schreiner2025}.

This large magnitude of the gravitational effect makes a measurement of the gravitational shift of the WGS experimentally accessible. 

It is enough to simply compare the results of the measurement with the reflective surface of the cylindrical mirror directed (almost) downward with the results of the measurement with the reflective surface of the cylindrical mirror (almost) upward. The remaining parameters of the experiments are identical.

Moreover, the current absolute precision of the theoretical description of interference patterns is at least $\sim 10^{-2}-10^{-3}$, and it can be improved. Therefore, even a comparison of the theoretical and experimental patterns provides information about the gravitational effect. 

\subsection{Simulation of an \Hydrogen{} WGS interference pattern}

The optimal velocity of \Hydrogen{} in Eq. (\ref{eq: hydrogen WGS velocity}) is close to the characteristic velocities of the \Hydrogen{} beam in the GQS experiment performed within the framework of the GRASIAN project \cite{Killian2023,Killian2024}. So we simply use the optimum parameters to simulate the interference pattern that we would like to measure: 
\begin{equation}
L_{WGS}^H=0.3~\text{m} \quad ; \quad V_{WGS}^{H}=1.2\cdot 10^2~\text{m/s} \quad ; \quad R_{WGS}^{H}=1.6\cdot 10^2~\text{m} \quad .
\end{equation}

The efficiency of the critical energy of the mirror as a "filter" of atomic WGS is limited because the lifetimes of WGS with energy above the critical energy do not decrease sharply. Therefore, we will add one more element to the experimental setup: an additional scatterer/absorber in the initial part of the mirror. By changing the height of the scatterer/absorber above the mirror, it is possible to change the number of WGS that can pass through the slit with small losses. We are not interested in an "absolutely sharp" cut-off of the spectrum of WGS. Therefore, here and in all similar cases below, we will choose the absorber/scatterer length short and equal to the length of a quasi-classical bounce, or $L_{WGS}/\beta_{WGS}$.

\begin{figure}[ht!]
    \centering
     \includegraphics[width=15cm] {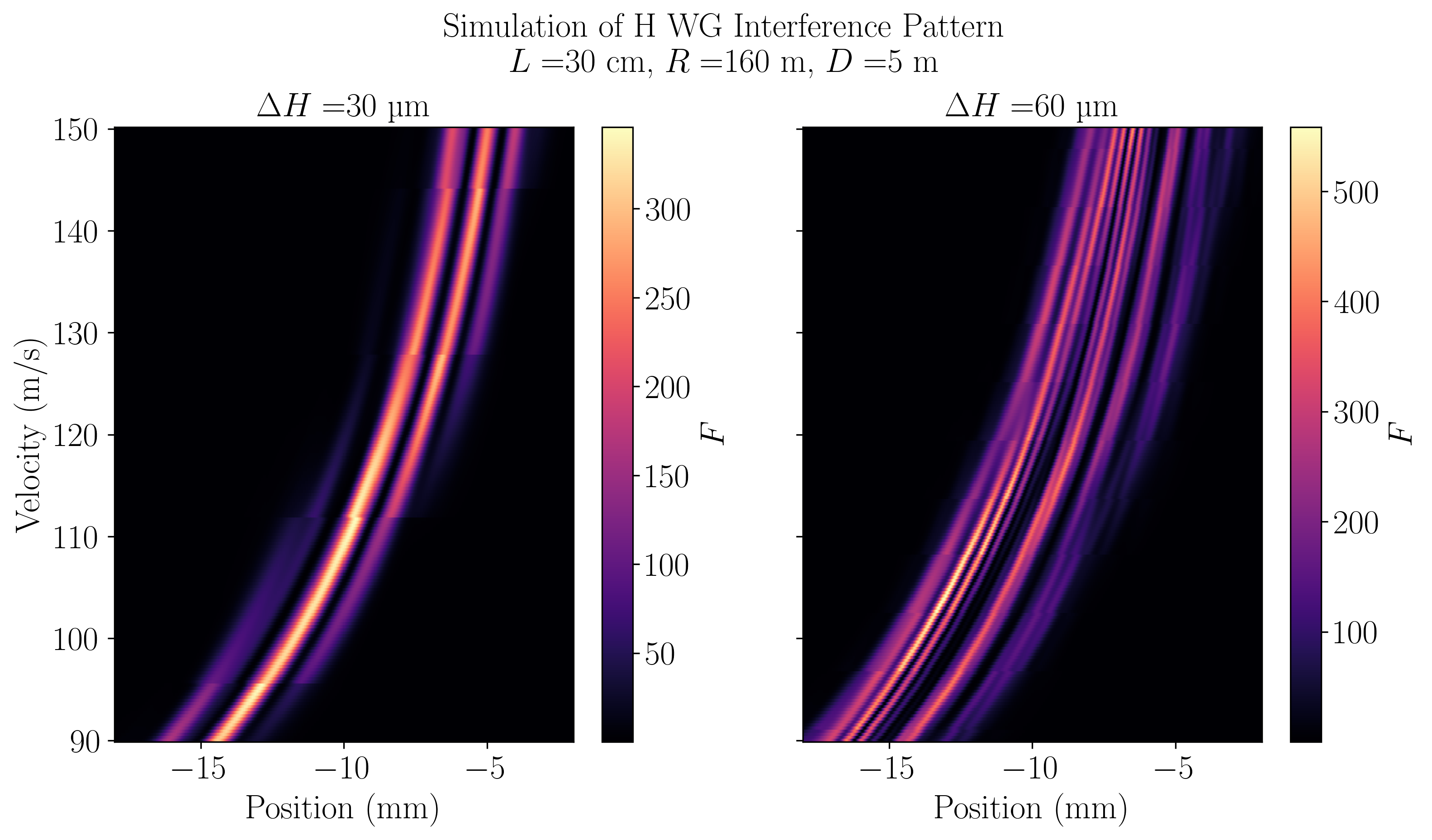}
  \caption{\label{fig:H_WG_Falling} Simulations of the expected interference pattern for the cylindrical mirror with a radius $R_{WGS}^H = 160$~m and a length $L_{WGS}^H = 30$~cm. Two scenarios are shown, each with an absorber/scatterer of length $(L_A)_{WGS}^H= 3$~cm placed at a height $\Delta H_{WGS}=30$~$\mu$m and $\Delta H_{WGS}=60$~$\mu$m above the beginning of the mirror. An incident plane wave is assumed to enter the gallery with no incidence angle and is propagated through the gallery by solving the Schrödinger equation. The wave packet at the exit of the gallery is then propagated through a free fall region of length $D_{WGS}^H = 5$~m. This image does not consider the velocity distribution of the \Hydrogen{} beam, nor the resolution effects of the detection system. The gravitational acceleration points downwards, and the origin corresponds to the edge at the exit of the gallery.}
\end{figure}

To resolve the peaks seen in the interference pattern on the left of Fig. \ref{fig:H_WG_Falling} and the most prominent peaks of the more intricate pattern seen on the right of Fig. \ref{fig:H_WG_Falling}, the chopper time window should not exceed $\sim3$~ms, and the spatial resolution of the position sensitive detector should not be worse than $\sim 300$~$\mu$m.
Ultimately, what determines the visibility of the interference structure is the velocity resolution of the hydrogen atoms. In the GRASIAN experiment, this resolution is defined by the temporal width of the beam pulses produced by the mechanical chopper in the cryogenic hydrogen beamline. The chopper selects narrow time slices of the continuous atomic flux, effectively constraining the longitudinal velocity spread through the relation $\Delta v / v \approx \Delta t / t_{\mathrm{TOF}}$, where $t_{\mathrm{TOF}}$ is the mean time of flight to the detection region. For typical flight times of a few tens of milliseconds, a chopper window of about $3$~ms corresponds to a relative velocity resolution of the order of $10^{-2}-10^{-1}$. Further improvements in the chopper timing or in the detection distance would directly translate into enhanced resolution of the higher-order interference features.
To resolve the fine structure of this interference pattern, the tighter constraints of a $\sim 0.5$~ms time window and $\sim 50$~$\mu$m position resolution are needed. In the following, we will give more details about how the measurement will be performed simply because it is already planned for the near future.

\subsection{A scheme of an \Hydrogen{} WGS experiment}

\begin{figure}[ht!] 
    \centering
     \includegraphics[width=10cm]{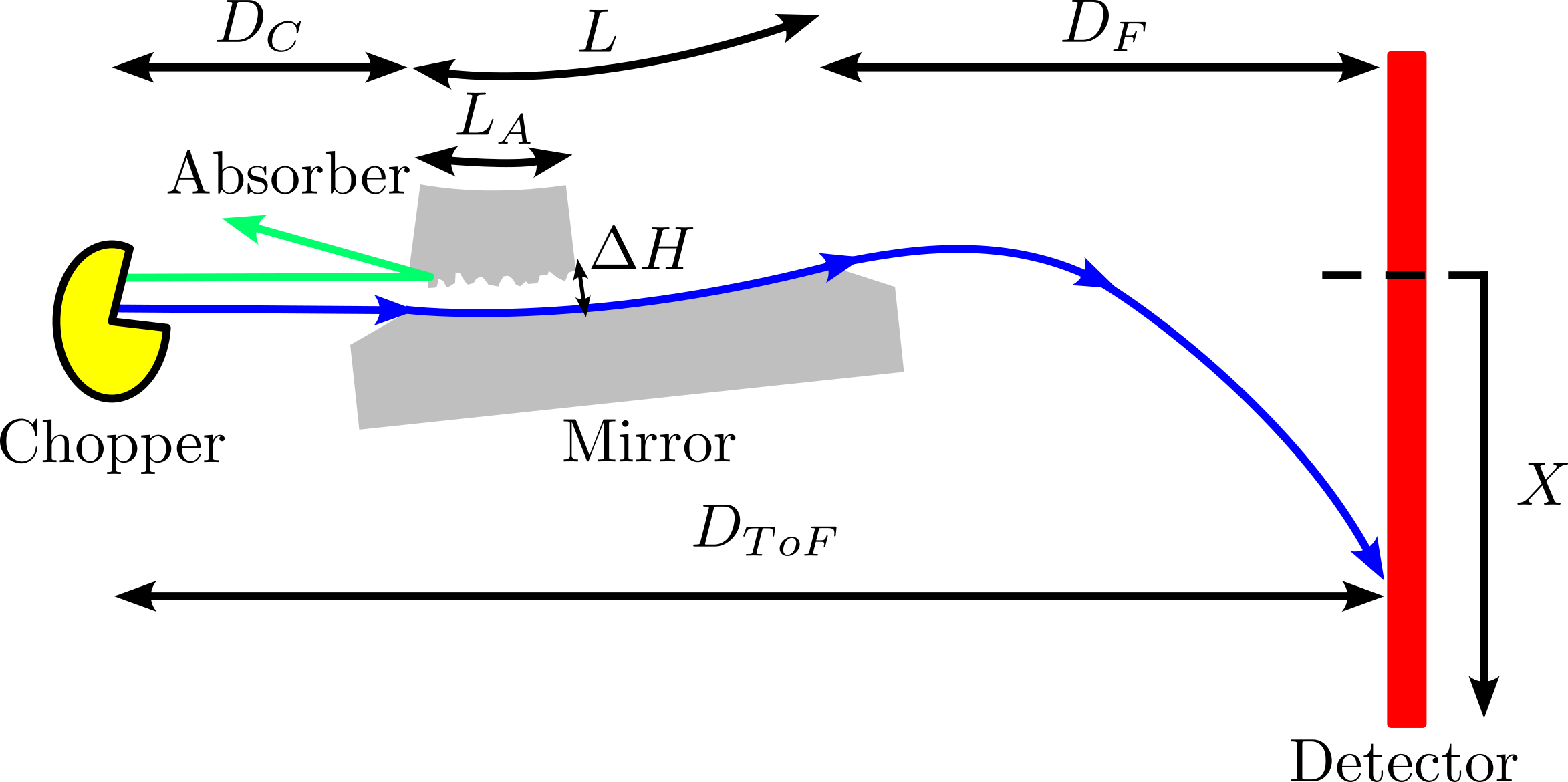}
    \caption{A schematic of the \Hydrogen{} WGS experiment. A collimated beam of \Hydrogen{} enters the system through a chopper, placed a distance $D_C$ away from the entrance of the mirror, which enables the measurement of the velocity of \Hydrogen{} atoms by the time of flight (ToF) method. The beam then enters the mirror absorber/scatterer system. Lower energy WGS can pass through the initial section where an absorber/scatterer of length $L_A$ is placed a height $\Delta H$ above the mirror, while higher WGS are rejected. The lower WGS then propagate along the surface and interfere with each other until reaching the end of the mirror, which has a total length $L$. The exiting beam enters a free fall region where a time and position sensitive detection system is placed a distance $D_F$ from the exit of the mirror. The position of the \Hydrogen{} atom on the detector is denoted as $X$.}
    \label{fig: WGS Absorber Interference}
\end{figure}

To facilitate the measurement of a gravitational shift $g/a_{WGS}^H$, we maximize it. For our cylindrical mirror with a very large radius of curvature, this means that first we measure with the cylindrically polished side facing upward, then with it facing downward, and compare the interference patterns. A complementary method is to compare the experimental and theoretical patterns. As the gravitational shift is large ($1.1\cdot10^{-1}$, Eq. (\ref{eq: hydrogen gravitational shift})) both methods are easy to implement; for the comparison of interference patterns with opposite orientation, the effect is twice larger.

This measurement allows prototyping experiments with \Antihydrogen{} and evaluation of the sensitivity to gravitational shift as well as the accuracy of the estimation of this value.

\subsection{A cryogenic beam of slow velocity \Hydrogen{} for WGS measurements}

The measurement of the WGS of \Hydrogen{} will take place at the Marietta Blau Institute (MBI) in Vienna with the cold \Hydrogen{} beamline developed at ETH Zurich in recent years \cite{Killian2023,Killian2024}. The setup is being prepared for the observation of GQS at the time of writing this article. 

The beam of atomic \Hydrogen{} is generated by dissociating H$_2$ molecules in a microwave cavity and directing the \Hydrogen{} atoms with a Teflon tube to a copper nozzle attached to the cold head in an optimized geometry \cite{10.1063/1.5129156}. By collisions with the cold nozzle, \Hydrogen{} is cooled to $\sim 6$~K. A chopper allows one to determine the longitudinal velocity of the beam by time-of-flight measurement, selecting the slow tail of the velocity distribution \cite{Killian2023,Killian2024}. 

In parallel, a new \Hydrogen{} source operating at the temperature of $\sim100$~mK has been developed in Turku by members of the GRASIAN collaboration ~\cite{Ahokas2022,Semakin2025}. Such a source would  increase the number of atoms with velocities below $100~\mathrm{m/s}$ by at least 3 orders of magnitude, substantially enhancing the fraction of atoms suitable for WGS measurement and opening the way to higher-contrast interference and improved statistics in future experiments. Further increase in phase space density could be achieved via magnetic trapping and cooling of the atoms.

\subsection{Resolution of the \Hydrogen{} WGS interference pattern}

The time resolution is defined by the opening time of the chopper, which is $\sim 5$~ms at the moment, and to a small extent also by the duration of the laser pulse. The spatial resolution of $\sim 120$~$\mu$m of the current detection scheme of \Hydrogen{} \cite{Killian2023,Killian2024} is defined by the width of the laser beam at the point of detection. 

A new chopper will be used, with an opening time of 0.5-3.0~ms, improving the resolution; the actual value should match the spatial resolution obtained. Improvement of the spatial resolution is more complicated, however, is possible using alternative methods, for example, with microchannel plates.

\subsection{A demonstration of the GQS of \Hydrogen{} using an interference method}

\begin{figure}[ht!]
    \centering
     \includegraphics[width=10cm]{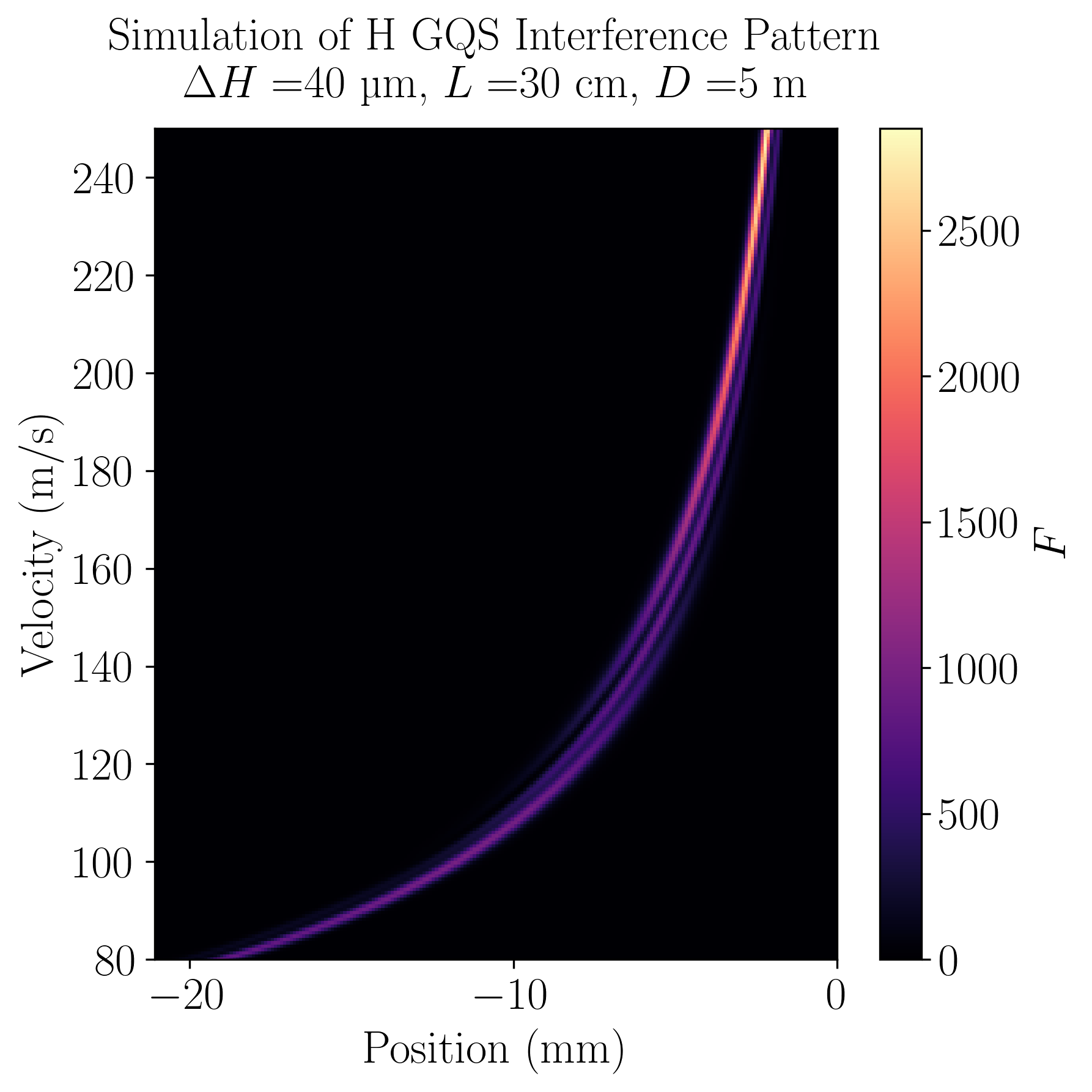}
    \caption{A simulation of the interference pattern generated by \Hydrogen{} propagating through a mirror-absorber/scatterer system of the same type as proposed for the VCN GQS measurement. The length of the system is $L_{GQS}^{H} = 30~\text{cm}$ and the absorber/scatterer height is $\Delta H_{GQS}^{H}=40~\mu
    \text{m}$. \Hydrogen{} enters a free fall region of length $D_{GQS}^{H} =5~\text{m}$ after leaving the mirror. Note that, just as it was in the case of proposed experiments with VCN, a much richer interference pattern can be observed if the detector resolution is higher, the distance is larger or/and the velocity is lower.}
    \label{fig: H falling}
\end{figure}

Using the installation as described in the previous sections and the method as described in Section \ref{sec: GQS with VCN} for VCNs, one can prove the existence of GQS of \Hydrogen{}. The only slight difference consists of a flight path available ($\sim 5$~m), a spatial resolution of the detector ($\sim 120$~$\mu$m), and a bit different velocity range (see Fig. \ref{fig: H falling}).

\section{Measurement of the WGS of antihydrogen and its gravitational shift}
\label{sec: experiment with antiH}

The WGS and GQS of \Antihydrogen{} are of interest for an investigation of the gravitational interaction of antimatter \cite{Nesvizhevsky2002,Dufour2014,Nesvizhevsky2019,Crepin2019}. The argument can be made that the gravitational interaction between matter and antimatter is well known from tests of the equivalence principle with objects that contain antimatter, i.e., nucleons. Kostelecky et al. \cite{Kostelecky2011} discuss how the constraints from these experiments can be evaded. The first result of a gravitational interaction test with \Antihydrogen{} has recently been achieved in \cite{Anderson2023}. Alternatively, the experimental accuracy of this comparison should be high enough to evade the existing indirect constraints. 

\subsection{Optimization of experiments with the WGS of \Antihydrogen}

A specific feature of experiments with \Antihydrogen{} is the difficulty in producing them sufficiently cold. Therefore, we will use Silica for the mirror, a material with a high effective critical energy, as we did in the case of \Hydrogen. An even larger effective critical energy for the mirror material would be advantageous because, on the one hand, it would expand the range of acceptable \Antihydrogen{} velocities and, on the other hand, it would increase the number of interference fringes, thereby increasing the accuracy of gravitational shift measurements. Since the list of materials for which this value is known is very limited, it is highly likely that such materials exist and could be used to fabricate a mirror. Therefore, searching for such materials is an important task. For now, we use Silica and evaluate the maximum values $V_{WGS}^{\bar{H}}$, at which the experiment is still feasible.

As the acceptable velocity of \Antihydrogen{} increases with increasing mirror length, we set it to a high value \footnote{Some increase in the mirror length is possible, and this option should be carefully analyzed if the proposed method is to be considered for a specific project.}:
\begin{equation}
\label{eq: antihydrogen mirror length}
  L_{WGS}^{\bar{H}}\sim 1~\text{m} \quad .  
\end{equation}

The acceptable velocity of \Antihydrogen{} also increases with increasing mirror radius $R_{WGS}^{\bar{H}}$. However, a precision mirror with a too large radius of curvature is difficult to manufacture, and the curvature of the mirror surface would not be maintained with sufficient accuracy. Moreover, it cannot be too large to avoid a direct view through the slit between the mirror and the scatterer/absorber \footnote{Strictly speaking, there is no direct view in the framework of purely quantum description but this quasi-classical model is still useful for the preliminary estimation of the experimental parameters}. To reduce this effect, the length of the absorber/scatterer must be increased when observing a large number of quantum states. All of these factors are difficult to quantify simultaneously, so we rely on our previous experience with mirrors and take for the maximum realistic radius of the mirror curvature the value of 
\begin{equation}
    \label{eq: antihydrogen mirror radius}
    R_{WGS}^{\bar{H}}\sim 10^4~\text{m} \quad .
\end{equation}

To increase the effective critical energy, we prefer a limited number of quasi-classical bounces. A comfortable number of quasi-classical bounces is, say:
\begin{equation}
    \label{eq: antihydrigen number of bounces}
    \beta_{WGS}^{\bar{H}}\sim 5 \quad .
\end{equation}
However, we will simulate the interference pattern up to maximum values of $V_{WGS}^{\bar{H}}$ thus exploring also the regime of $\beta_{WGS}^{\bar{H}}\sim 1$.

To observe an interference pattern with a significant number of interference lines, the system should provide at least a few WGS:
\begin{equation}
    \label{eq: antihydrigen number of states}
  \gamma_{WGS}^{\bar{H}}\sim 3+2 \quad .
\end{equation}

This combination of parameters results in the following set of effective critical energy $E_{lim}^{\bar{H}}$, acceleration $a_{WGS}^{\bar{H}}$, and velocity $V_{WGS}^{\bar{H}}$ for an experiment with \Antihydrogen:
\begin{equation} 
    \label{eq: parameters for antihydrogen}
E_{lim}^{\bar{H}}\sim 2.0\cdot 10^{-10}~\text{eV}; \quad
 a_{WGS}^{\bar{H}}\sim 5.4\cdot 10^3~\text{m/s$^2$}; \quad V_{WGS}^{\bar{H}}\sim 7.4\cdot 10^3~\text{m/s} \quad .    
\end{equation}

The measurement can be performed with \Antihydrogen{} of thermal velocities, or several times larger (we will evaluate in Section \ref{sec: simulation for antihydrogen} up to what velocity of \Antihydrogen{} the range can be, in principle, extended). For an optimal velocity of $V_{WGS}^{\bar{H}}\sim 7.4\cdot 10^3$~m/s, the gravitational shift is $\sim 1.8\cdot 10^{-3}$ which can be easily measured. Smaller velocities of the \Antihydrogen{} would be an advantage. Higher velocities of \Antihydrogen{} can also be considered; however, the gravitational shift decreases quadratically with increasing velocity. Fabrication of required mirrors is feasible.

In order to avoid extremely precise absolute measurements, it seems reasonable to measure the WGS of \Antihydrogen{} and \Hydrogen{} using the same mirror and compare the results. Theoretical corrections responsible for the small difference in QR of \Antihydrogen{} and \Hydrogen{} should then be introduced.

\subsection{A scheme of an \Antihydrogen{} WGS experiment}
\label{sec: scheme for antihydrogen}

To propose an experimental setup for \Antihydrogen, one should keep the following factors in mind: the angular spread of \Antihydrogen{} at the mirror exit is very small due to the very large longitudinal velocities of \Antihydrogen{} and the very low transverse velocities. Therefore, the experimental setup possible with \Neutron{} and \Hydrogen{} (Fig. \ref{fig: WGS Absorber Interference}) is not applicable directly here. 

However, another option is possible. When touching the mirror surface, the \Antihydrogen{} is annihilated and the signal from such an event (charged pions) can be registered in a position-sensitive detector. For this reason, the "useful signal" that we simulate in Section \ref{sec: simulation for antihydrogen} is the flow of \Antihydrogen{} into the mirror surface. Another difference is that the longitudinal velocity of the \Antihydrogen{} atom is not set by the chopper; the spectrum is monochromatic or is set by the moment of release of \Antihydrogen{} from the trap (depending on the specific implementation of the experiment). 

\begin{figure}[ht!]
    \centering
     \includegraphics[width=8cm]{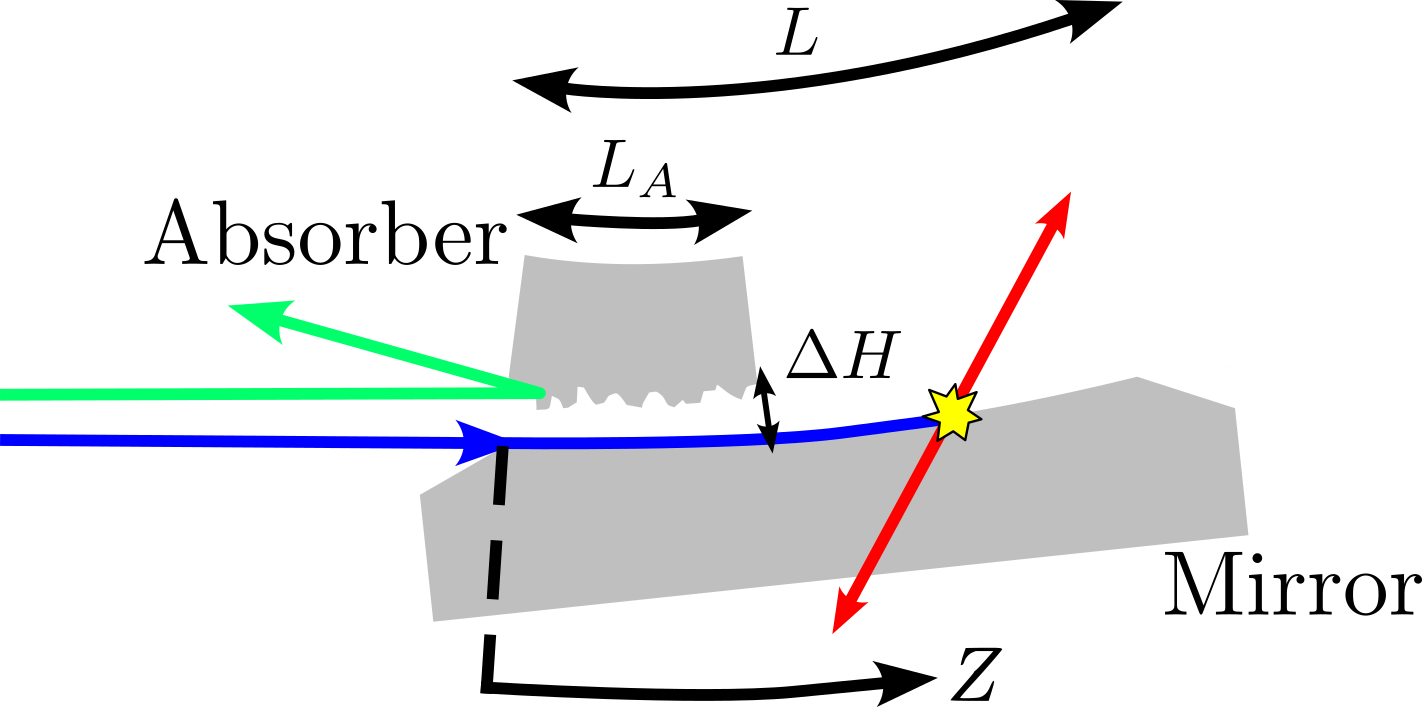}
    \caption{A schematic of the \Antihydrogen{} WGS experiment. A collimated monochromatic beam of \Antihydrogen{} enters the mirror absorber/scatterer system. Like with the \Hydrogen{} WGS set-up before, an absorber/scatterer of length $L_A$ is placed a height $\Delta H$ above the mirror and filters out higher energy WGS. The surviving WGS propagate along the surface and interfere over the mirror of length $L$. The \Antihydrogen{} atoms which are not reflected by the Casimir-Polder interaction will reach the surface and annihilate. The position $Z$ of this annihilation event is taken as the signal and is experimentally determined by recording the trajectories of the annihilation products. The predicted annihilation rate can be taken as the probability current of the \Antihydrogen{} wave function at the mirror surface,}
    \label{fig:WG Current with Absorber}
\end{figure}

The last "free parameter" in this experimental setup (Fig. \ref{fig:WG Current with Absorber}) is the height $\Delta H_{WGS}^{\bar{H}}$ of the absorber/scatterer above the mirror. By increasing $\Delta H_{WGS}^{\bar{H}}$, we can increase the number of observed interference fringes and thus increase the sensitivity of the experiment to measuring gravitational acceleration. However, this increase is limited by the spatial resolution of the position-sensitive detector of \Antihydrogen{} annihilation, the direct view through the slit mirror-absorber/scatterer as well as by the decrease in the lifetime of the WGS with increasing quantum state number.

\subsection{Simulation of an \Antihydrogen{} WGS interference pattern}
\label{sec: simulation for antihydrogen}

The parameters used for the simulation are:
\begin{equation}
    \label{eq: antihydrogen set of parameters}
    L_{WGS}^{\bar{H}}=1~\text{m}; \quad  R_{WGS}^{\bar{H}}=10^4~\text{m}; \quad V_{WGS}^{\bar{H}}=7.4\cdot 10^3~\text{m/s} \quad .
\end{equation}

A characteristic state size $l_{WGS}^{\bar{H}}$ is:
\begin{equation}
    \label{eq: antihydrogen characteristic slit size}
    l_{WGS}^{\bar{H}}\sim 0.71~\text{$\mu$m} \quad .
\end{equation}

To allow for N excited WGS, one should set the slit to the height:
\begin{equation}
    \label{eq: antihydrogen slit size}
    \Delta H_{WGS}^{\bar{H}}\sim l_{WGS}^{\bar{H}}\cdot (3+N) \quad .
\end{equation}

\begin{figure}[ht!]
    \centering
     \includegraphics[width=10cm]{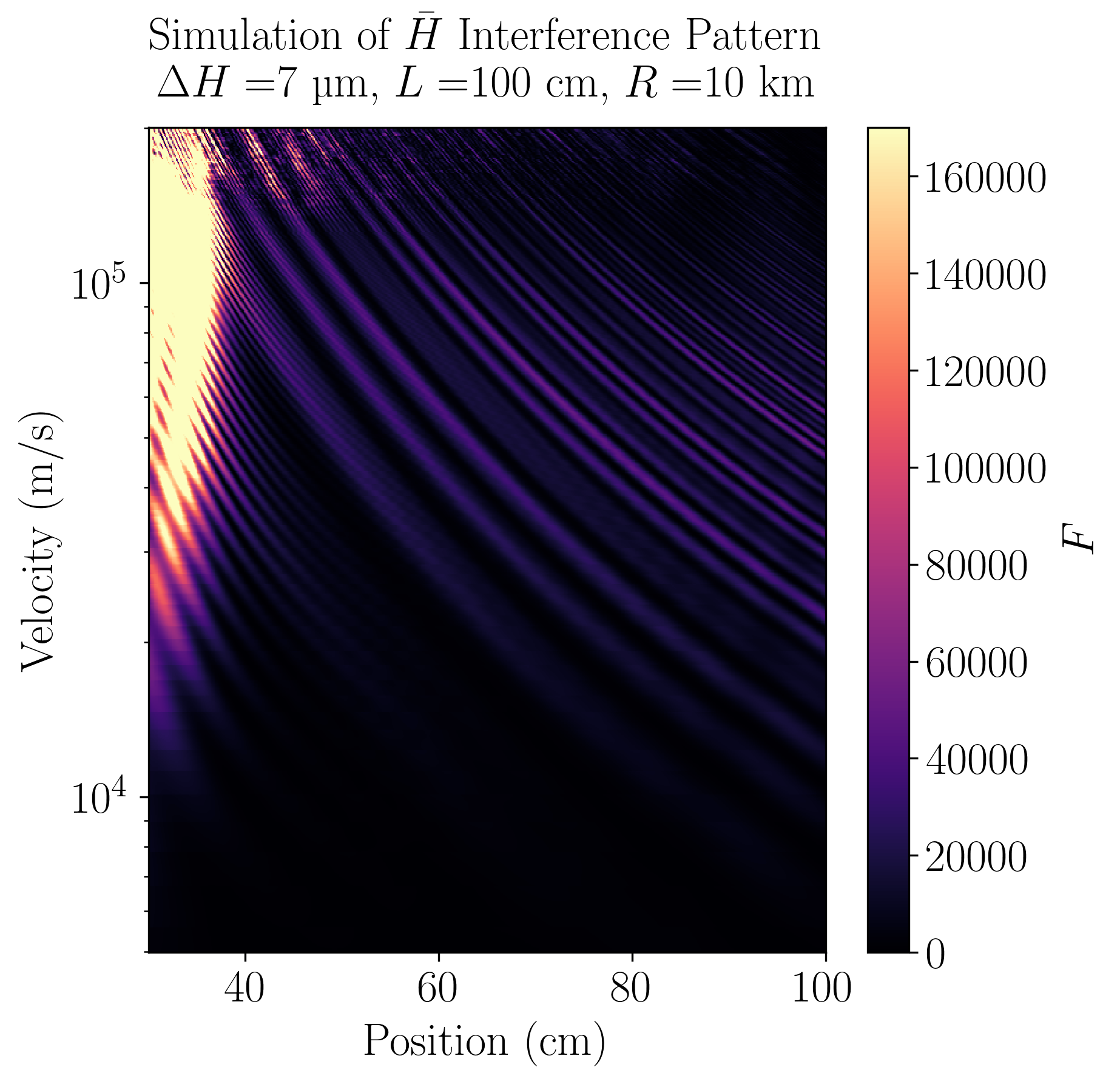}
\label{fig:Hbar_WG_Current}
\caption{A simulation of the annihilation rate of $\bar{H}$ on the surface of a WGS mirror with the parameters in Eq. (\ref{eq: antihydrogen set of parameters}) and a wide range of velocities. An absorber/scatterer at a height $\Delta H_{WGS}^{\bar{H}}\sim 7~\text{$\mu$m}$ allows 106 states to be populated for $200.000$~m/s and 4 states for $5000$~m/s in the absorber/scatterer region. The length of the absorber/scatterer region is $(L_a)_{WGS}^{\bar{H}} = 30$~cm.  The spatial resolution of the \Antihydrogen{} annihilation detector needed to measure WGS with the \Antihydrogen{} velocity of $200.000$~m/s is $\sim 1$~mm, and it is $\sim 5$~mm for the \Antihydrogen{} velocity of $5000$~m/s.}
    \label{fig: Hbar Current}
\end{figure}

The interference pattern simulated in Fig. \ref{fig: Hbar Current}, can be measured with velocities of \Antihydrogen{} up to $\sim 10^{5}$~m/s. They are an order of magnitude lower than those available in the GBAR experiment \cite{Adrich2023}. Moreover, the loss of \Antihydrogen{} atoms increases rapidly with increasing velocity. The flux of \Antihydrogen{} decreases by an order of magnitude if the velocity of \Antihydrogen{} increases by a factor of $\sim 2$. "Fine tuning" of the experimental parameters ($L_{WGS}^{\bar{H}}$, $R_{WGS}^{\bar{H}}$, $\Delta H_{WGS}^{\bar{H}}$, mirror material, etc.) for specific beam parameters is possible. However, smaller velocities of \Antihydrogen{} are always an advantage.

In principle, it is possible to perform a similar experiment even with \Antihydrogen{} velocities of $\sim 10^6$~m/s. This requires using the material with the highest critical energy, liquid helium. Since coating the surface of a cylindrical mirror with a thick layer of helium with stable, well-controlled, and consistent properties seems unrealistic, the mirror surface must be flat. To achieve the GQS/WGS effect, an additional force must be added, pressing the \Antihydrogen{} to the helium surface due to a magnetic field gradient. We may optimize this possibility if it is of practical interest. Unfortunately, it will not be an elegant, simple solution like most of the proposals in this article. In addition, the gravitational shift decreases quadratically and becomes too small to be measured reliably.

\section{GQS of Hydrogen and Antihydrogen on a "falling mirror"}

As noted above, the storage time of \Hydrogen{} and \Antihydrogen{} on the plane, due to QR from the surface, is limited to $\sim 1$~s. This is much longer than the observation times of the UCN in existing experiments, which can dramatically increase the accuracy of quantum gravitational spectroscopy and interferometry. Because the atoms stored in GQS interact with the surface in a well-known way, this interaction can be taken into account very accurately. This circumstance opens up the potential for Doppler-free optical and hyperfine spectroscopy of such atoms, since their effective velocities in the direction perpendicular to the plane (the velocity distribution in the lowest GQS) are record low.

It is of interest to consider the possibility of even longer storage of atoms on the surface and even more precise spectroscopy and interferometry. This can be achieved by placing the atoms in reduced gravity. In the laboratory, this effect can be achieved most simply by tilting the magneto-gravitational trap (MGT) \cite{Nesvizhevsky2020mgt}. However, it will be the same when the mirror falls with almost gravitational acceleration and when the experiment is actually placed in reduced gravity. In all these cases, we will simply consider the reduced gravitational acceleration $g_{GQS^{reduced}}^{(H,\bar{H})}$.

To understand the effect of reduced gravity, let us reproduce the dependences of all parameters on the value of $g_{GQS^{reduced}}^{(H,\bar{H})}$:
\begin{equation}
  E_{GQS^{reduced}}^{(H,\bar{H})}\propto(g_{GQS^{reduced}}^{(H,\bar{H})})^{\frac{2}{3}}, l_{GQS^{reduced}}^{(H,\bar{H})}\propto (g_{GQS^{reduced}}^{(H,\bar{H})})^{-\frac{1}{3}}, \tau_{GQS^{reduced}}^{(H,\bar{H})}\propto (g_{GQS^{reduced}}^{(H,\bar{H})})^{-\frac{2}{3}}.
\end{equation}

The lifetime $t_{GQS^{reduced}}^{(H,\bar{H})}$ of \Hydrogen{} and \Antihydrogen{} in GQS, due to QR from the surface, is proportional to
\begin{equation}
  t_{GQS^{reduced}}^{H,\bar{H}}\propto (g_{GQS^{reduced}}^{(H,\bar{H})})^{-\frac{4}{3}} \quad . 
\end{equation}

The significant increase in storage time is easy to understand in quasi-classical approximation: with a decrease in gravity, it increases both by reducing the frequency of bounces on the surface and by increasing the wavelength of the atom.

At the same time, the precision measurement of the spectrum of GQS (in reduced gravity) $\Delta E_{GQS^{reduced}}^{(H,\bar{H})}/E_{GQS^{reduced}}^{(H,\bar{H})}$, which in a rough approximation is proportional to the reversal number of quasi-classical bounces, quickly improves:
\begin{equation}
 \frac{\Delta E_{GQS^{reduced}}^{(H,\bar{H})}}{E_{GQS^{reduced}}^{(H,\bar{H})}}\propto \frac{\tau_{GQS^{reduced}}^{H,\bar{H}}}{t_{GQS^{reduced}}^{H,\bar{H}}}\propto  (g_{GQS^{reduced}}^{(H,\bar{H})})^{-\frac{2}{3}} \quad . 
\end{equation}

In addition to the proven significant increase in the accuracy and sensitivity of spectroscopic and interferometric measurements, the requirements for optical elements in such experiments are greatly simplified. Since the characteristic sizes of quantum states become much larger, the requirements for surface roughness, the relative accuracy of mirror and absorber/scatterer installation, the resolution of position-sensitive detectors, etc. are significantly lower.

\section{WGS of Muonium}
\label{sec: muonium}

\subsection{Optimization of an experiment with the WGS of Muonium}

We turn to a discussion of the WGS of Muonium (\Muonium). Kostelecky et al. \cite{Kostelecky2011} discuss a non-standard Hamiltonian with new terms that depend on particle species. Investigations of the new coupling constants to higher generation particles are rare and of independent interest. The authors of ref. \cite{Kostelecky2011} expect that the new coupling constants increase with the fermion mass.

When considering the feasibility of measuring the WGS of \Muonium, there are several limitations and advantages that should be taken into account. 

On the "limitations" side, the \Muonium{} lifetime is short: 
\begin{equation}
\label{eq: muonium lifetime}
\tau^{Mu}\sim 2.2\cdot 10^{-6}~\text{s} \quad ,   
\end{equation}
therefore, the observation time of the WGS of \Muonium{} ($\tau^{Mu}_{WGS}$) cannot exceed this value substantially.

However, the use of sources of \Muonium{} at cryogenic temperatures \cite{Antognini:2011ei} or the novel beam of \Muonium{} produced from superfluid helium \cite{Zhang:2025ogj} with a velocity of 
\begin{equation}
\label{eq: muonium velocity}
  V_{WGS}^{Mu}\sim 2.2\cdot 10^3~\text{m/s}  
\end{equation}
combined with the planned High Intensity Muon Beamline (HIMB) upgrade at PSI \cite{HIMB_Status_2023} could allow the observation of WGS. 
 This velocity is even significantly lower than that of \Antihydrogen{} discussed in Section \ref{sec: experiment with antiH}. 

Due to the small mass of \Muonium,
\begin{equation}
\label{eq: muonium mass}
  m^{Mu}\sim 0.113\cdot m^{n,H,\bar{H}} \quad ,   
\end{equation}
the effective critical energy of the materials for \Muonium{} is much larger than their effective critical energy for the \Hydrogen{} and \Antihydrogen{} atoms. 

On the one hand, large effective critical energies simplify the observation of the WGS of \Muonium{}, and relax the constraints on \Muonium{} velocities. This is particularly useful if the goal is to investigate the QR itself. However, the relative value of the gravitational shift is much smaller, and the flux of \Muonium{} into the surface is suppressed. We will aim to optimize the experimental parameters to simplify the observation of the gravitational shift.  

The \Muonium{} emitted from superfluid helium \cite{osti_6584671, Soter_2021_SFHe} can be considered monochromatic (we will illustrate this statement quantitatively in Section \ref{sec: simulations for muonium}):  
\begin{equation}
\delta V^{Mu}\sim 5\cdot 10^1~\text{m/s} \quad ,  
\label{eq: monochromatic muonium}
\end{equation}
and the phase space density of the \Muonium{} beam is quite large.

For  
\begin{equation}
 \tau^{Mu}_{WGS}=3\cdot \tau^{Mu} \quad ,   
\end{equation}
the mirror length \footnote{We accept that only a small fraction of the beam can reach the exit of the mirror but try to maximally increase the sensitivity using this parameter ($L_{WGS}^{\Muonium{}}$), and the mirror length is small anyway.} is
\begin{equation}
\label{eq: muonium WGS length}
    L_{WGS}^{Mu}\sim V_{WGS}^{Mu} \cdot \tau_{WGS}^{Mu}\sim 1.5\cdot 10^{-2}~\text{m} \quad . 
\end{equation}

We begin optimizing this experiment by comparing it with the measurement of the magnetic shift of the neutron WGS \cite{Schreiner2025}. In this short measurement, the relative contribution of an additional force was $\sim 10^{-5}$. In a dedicated experiment, the sensitivity can greatly increase; however, we start optimizing the \Muonium{} experiment from this value. So, we get the first condition:
\begin{equation}
 a_{WGS}^{Mu}\sim 10^6~\text{m/s$^2$} \quad .   
\end{equation}

Thus, the characteristic time of formation of the WGS is (Eq. (\ref{eq: time mass acceleration},\ref{eq: muonium mass})):
\begin{equation}
\label{eq: muonium time formation}
    \tau_{WGS}^{Mu}\sim 9.3\cdot 10^{-7}~\text{s} \quad .
\end{equation}
Note that the formation time of the quantum state $\tau_{WGS}^{Mu}$ is shorter than the lifetime of \Muonium, $\tau^{Mu}$, so the number of quasi-classical bounces (Eq. \ref{eq: number of whispering quasiclassical bounces}) is small but acceptable:
\begin{equation}
\label{eq: muonium number of bounces}
\beta_{WGS}^{Mu}\sim \frac{\tau^{Mu}}{\tau_{WGS}^{Mu}}\sim 2.3 \quad .
\end{equation}

The mirror radius is comfortable for mirror manufacturing (Eqs. (\ref{eq: centrifugal acceleration},\ref{eq: muonium velocity},\ref{eq: muonium time formation})):
\begin{equation}
\label{eq: muonium mirror radius}
    R_{WGS}^{Mu}\sim 4.8~\text{m} \quad .
\end{equation}

The only parameter that is too large is the number of WGS. For instance, for $\beta_{WGS}^{Mu}$ (Eq. (\ref{eq: muonium number of bounces})) bounces on Silica, it is
\begin{equation}
\label{eq: muonium number of states}
   \gamma_{WGS}^{Mu}\sim 8.6\cdot 10^2 \quad ,
\end{equation}
and it is too large for any material.

Therefore, the choice of mirror material does not matter if we consider only the shape of the interference pattern \footnote{To increase the flux of \Muonium{} into the mirror surface, the effective critical energy should be low, thus, a conducting surface is a better choice.}, and we have to limit the number of WGS in a different way. As we did for \Hydrogen{} and \Antihydrogen{}, a short absorber/scatterer should be installed in the \Muonium{} experiment above the initial part of the mirror with a slit size $\Delta H_{WGS}^{Mu}$:
\begin{equation}
\label{eq: muonium slit size}
 \Delta H_{WGS}^{Mu}\sim l_{WGS}^{Mu}\cdot (3+N) \quad , 
\end{equation}
where 
\begin{equation}
  l_{WGS}^{Mu}\sim 5.6\cdot 10^{-7}~\text{m}  
\end{equation}
and $N$ is the number of excited WGS.

Although all the parameters of the problem are quite comfortable for observing the WGS of \Muonium{}, measuring the gravitational shift of the WGS of \Muonium{} is much more challenging. The relative shift is
\begin{equation}
    \frac{g}{a_{WGS}^{\Muonium{}}}\sim 10^{-5} \quad .
    \label{eq: gravitational shift muonium}
\end{equation}
This value (Eq. \ref{eq: gravitational shift muonium}) matches exactly the sensitivity of the short neutron experiment \cite{Schreiner2025}. Therefore, achieving the same sensitivity in the \Muonium{} experiment is not fundamentally difficult. However, unlike the neutron experiment, in which the direction of the additional force was reversed, in the case of measuring a gravitational shift of the WGS of \Muonium{}, one needs to reverse the orientation of the mirror. This means that one must ensure the mirror adjustment and control with corresponding accuracy, which is an extremely challenging task. In any case, prototyping this experiment with neutrons seems like a necessary preliminary step because all the parameters of this task are typical for neutron experiments, and the neutron fluxes are also high. For example, such an experiment can be performed at the PF1B facility at the ILL \cite{Abele2006}.

\subsection{Simulation of a \Muonium{} WGS interference pattern}
\label{sec: simulations for muonium}

Here, we summarize the experimental parameters for the WGS of \Muonium{} (Eqs. (\ref{eq: muonium velocity},\ref{eq: muonium WGS length},\ref{eq: muonium mirror radius},\ref{eq: muonium slit size})):
\begin{equation}
    \label{eq: summary of parameters for muonium}
V_{WGS}^{Mu}\sim 2.2\cdot 10^3~m/s ; L_{WGS}^{Mu}\sim 1.5\cdot 10^{-2}~\text{m} ; R_{WGS}^{Mu} \sim 4.8~\text{m} ; \Delta H_{WGS}^{Mu}\sim 5.6\cdot 10^{-7}\cdot (3+N)~\text{m} .
\end{equation}

\begin{figure}[ht!]
    \centering
     \includegraphics[width=10cm]{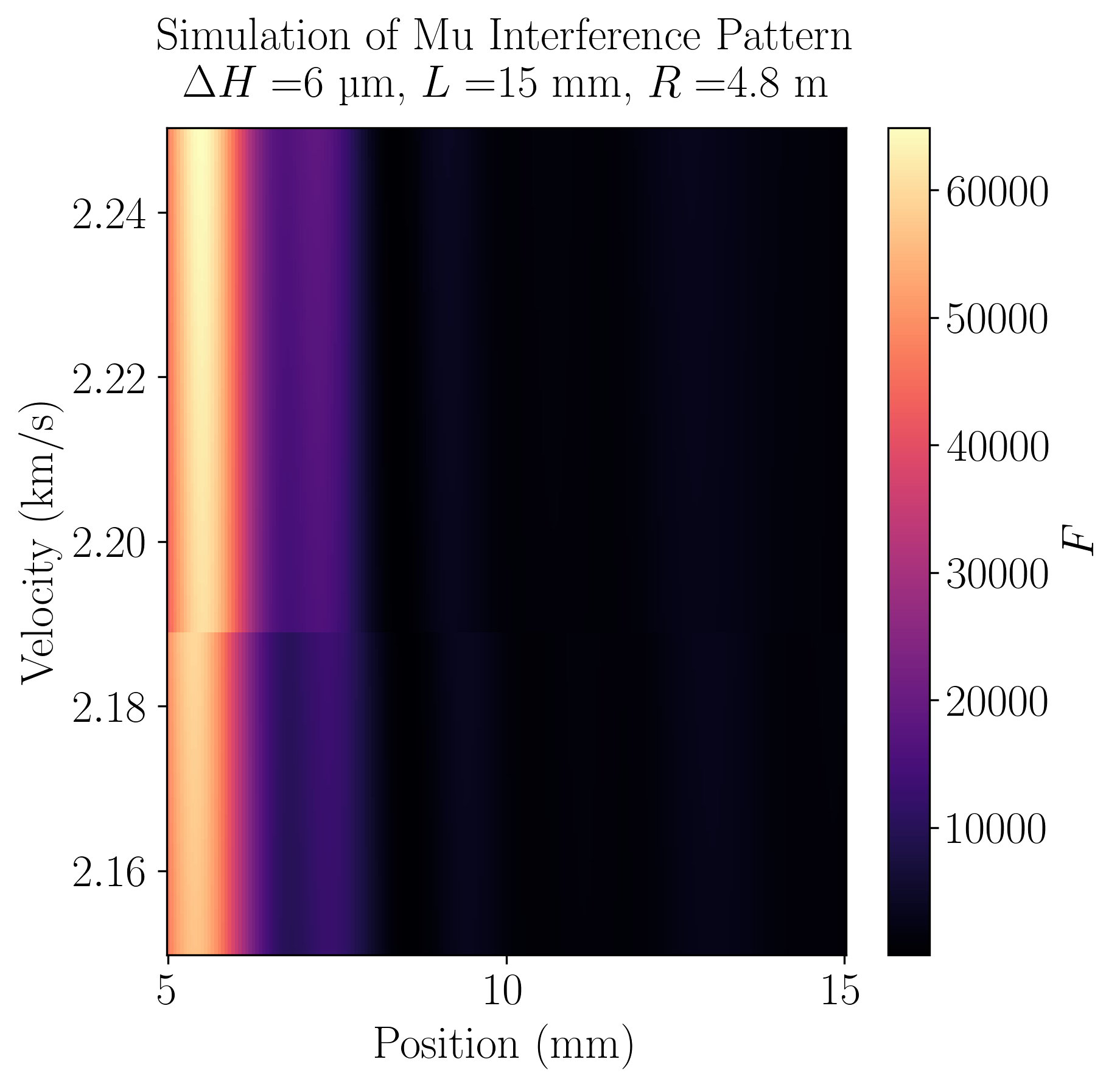}
\caption{A simulation of the \Muonium{} current into the surface of the WGS mirror with the parameters in Eq. \ref{eq: summary of parameters for muonium}. 8 states contribute to the pattern due to the presence of an absorber/scatterer of length $(L_A)_{WGS}^{Mu} = 5$~mm at the beginning of the mirror. The decay of \Muonium{} is taken into account. Only the region after the absorber/scatterer is presented since it's almost independent on precisely setting the $\Delta H_{WGS}^{Mu}$ value, thus more reliable to simulate.}
    \label{fig: Mu Current}
\end{figure}

\begin{figure}[ht!]
    \centering
     \includegraphics[width=10cm]{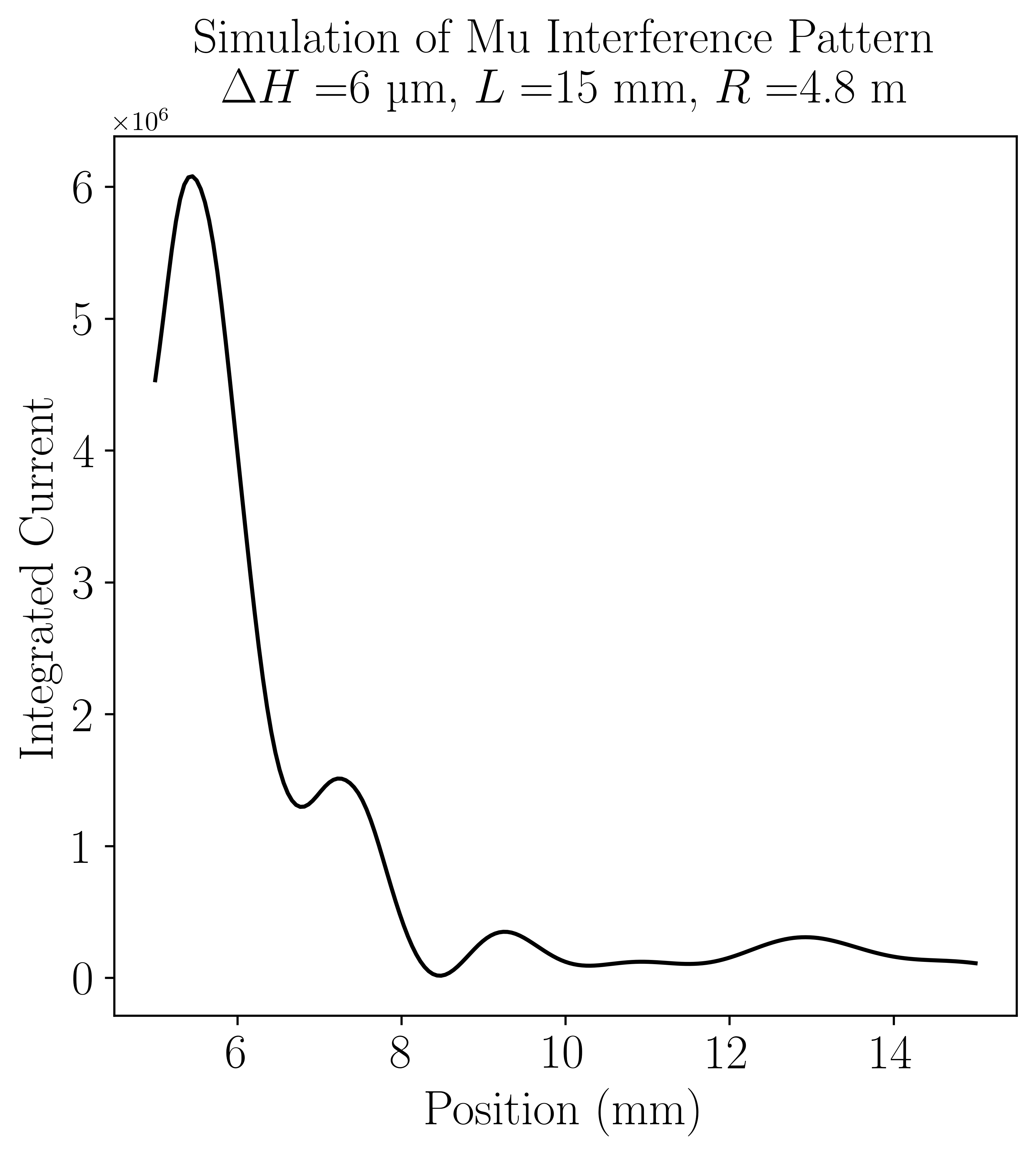}
\caption{The \Muonium{} current into the mirror calculated in Fig. \ref{fig: Mu Current} integrated over the expected velocity spectrum. The decay of \Muonium{} is taken into account. One can see that the assumption of a "monochromatic beam" (Eq. \ref{eq: monochromatic muonium}) is valid, and the interference contrast is very high.} 
    \label{fig: Mu Integrated Current}
\end{figure}

The simulation results are shown in Figs. \ref{fig: Mu Current} and \ref{fig: Mu Integrated Current}. Note that the purpose of this simulation was only to demonstrate the possibility of observing the WGS of \Muonium{}. To increase the chances of observing the gravitational shift of the WGS of \Muonium{}, it is preferable to use only excited WGS by tilting the incident beam relative to the mirror surface. In this case, the number of interference fringes that can be observed will increase greatly and the accuracy of the shift measurement will also improve greatly. However, this is a more complex optimization that is far beyond the scope of this article.

\subsection{A scheme of a \Muonium{} WGS experiment}

The scheme for WGS formation is similar to that used for \Hydrogen{} and \Antihydrogen. However, the detection scheme of \Muonium{} would differ from the first and second cases. Unlike \Hydrogen, both the lifetime of \Muonium{} is too short and the angular spread of \Muonium{} is too small to be analyzed using a position-sensitive detector installed a large distance from the edge of the mirror. Unlike \Antihydrogen, measuring the annihilation points of \Muonium{} in the mirror also does not work. If the mirror is magnetic, the spin rotation of \Muonium{} upon contact with the mirror may allow the WGS pattern to be detected. However, a detection method requires more serious analysis and is not clear to us at the moment.

\section{WGS of Positronium}
\label{sec: positronium}

When considering the possibility of measuring the WGS of \Positronium, several factors should be considered.

First, since the mass of \Positronium{} is much smaller than the mass of \Muonium, the probability of QR and the effective critical energy are even much higher. Therefore, we would use the same experimental scheme as in the case of \Muonium, i.e., excess excited WGS are removed using an absorber/scatterer. However, in order to increase the probability that \Positronium{} reaches the surface and annihilates, we send the \Positronium{} beam at a significant angle to the surface. This method allows us to eliminate the low WGS. As a result, the signal is stronger and the interference pattern contains more interference lines.

Second, the lifetime of \Positronium{} in the ground state is even shorter than the lifetime of \Muonium. However, the lifetime of the excited quantum state increases rapidly with increasing quantum number. This circumstance can be understood as follows: with an increasing quantum number, the average distance between the electron and positron increases, the overlap integral of their wave functions decreases, and, accordingly, the probability of annihilation decreases. For example, for the quantum state $n=33$, the lifetime is
\begin{equation}
    \tau^{Ps,n=33}\sim 10^{-5}~\text{s} \quad .
\end{equation}

A characteristic velocity of available \Positronium{} beams is \cite{PhysRevA.81.052703, PhysRevA.81.012715}
\begin{equation}
\label{eq: positronium velocity}
    V^{Ps}\sim 5\cdot 10^4~\text{m/s} \quad .
\end{equation}

Assuming that the time of flight of the \Positronium{} over the mirror ($t_{WGS}^{Ps,n=33}\sim L_{WGS}^{Ps,n=33}/V^{Ps}$) is equal to the lifetime $\tau^{Ps,n=33}$, the length of the mirror is
\begin{equation}
\label{eq: positronium mirror length}
    L_{WGS}^{Ps,n=33}\sim 0.5~\text{m} \quad .
\end{equation}

Based on these values, we will optimize the experimental design to increase the chances of observing a gravitational shift of \Positronium.

To decrease a relative gravitational shift, we prefer a small number of quasi-classical bounces:
\begin{equation}
    \beta_{WGS}^{Ps,n=33}\sim 3 \quad .
\end{equation}

Then the time of formation of the WGS of \Positronium{} is
\begin{equation}
    \tau_{WGS}^{Ps,n=33}\sim t_{WGS}^{Ps,n=33}/\beta_{WGS}^{Ps,n=33}\sim 3.3\cdot 10^{-6}~\text{s},
\end{equation}
and the acceleration is 
\begin{equation}
\label{eq: positronium acceleration}
    a_{WGS}^{Ps,n=33}=1.8\cdot 10^6~\text{m/s$^2$} \quad ,
\end{equation}
and the relative gravitational shift is 
\begin{equation}
\label{eq: positronium gravitational shift}
   \frac{g}{a_{WGS}^{Ps,n=33}}\sim 5.5\cdot 10^{-6} \quad . 
\end{equation}
Note that this value (\ref{eq: positronium gravitational shift}) is as large (as small) as the sensitivity $\sim 10^{-5}$ of the neutron test experiment \cite{Schreiner2025} or the estimated effect in the experiment with \Muonium{} (Eq. \ref{eq: gravitational shift muonium}).

The radius of the mirror $R_{WGS}^{Ps,n=33}$, which corresponds to the velocity (Eq. (\ref{eq: positronium velocity})) and acceleration (Eq. (\ref{eq: positronium acceleration})) of \Positronium, is equal to
\begin{equation}
    \label{eq: positronium radius}
    R_{WGS}^{Ps,n=33}\sim 1.4 \cdot 10^3~\text{m} \quad .
\end{equation}

One significant simplification of such an experiment is that the size of the slit to shape the WGS is relatively large due to the small mass of $Ps
$. The characteristic slit size is:
\begin{equation}
    \label{eq: positronium state characteristic size}
l_{WGS}^{Ps,n=33}\sim 9.8\cdot 10^{-6}~\text{m} \quad .
\end{equation}
For
\begin{equation}
    \gamma_{WGS}^{Ps,n=33}\sim 10 \quad ,
\end{equation}
it is
\begin{equation}
    \label{eq: positronium slit height}
\Delta H_{WGS}^{Ps,n=33}\sim 9.8\cdot 10^{-5}~\text{m} \quad .
\end{equation}

\subsection{Simulation of a \Positronium{} WGS interference pattern}

Let us summarize the characteristic parameters of the problem:
\begin{equation}
    n=33 ; V_{WGS}^{Ps,n=33}\sim 5\cdot 10^4~\text{m/s} ; L_{WGS}^{Ps,n=33}\sim 0.5~\text{m} ; g/a_{WGS}^{Ps,n=33}\sim 5.5\cdot 10^{-6} ; R_{WGS}^{Ps,n=33}=1.4\cdot 10^3~\text{m} .
    \label{eq:
    parameters for positronium}
\end{equation}

\begin{figure}[ht!]
    \centering
     \includegraphics[width=10cm]{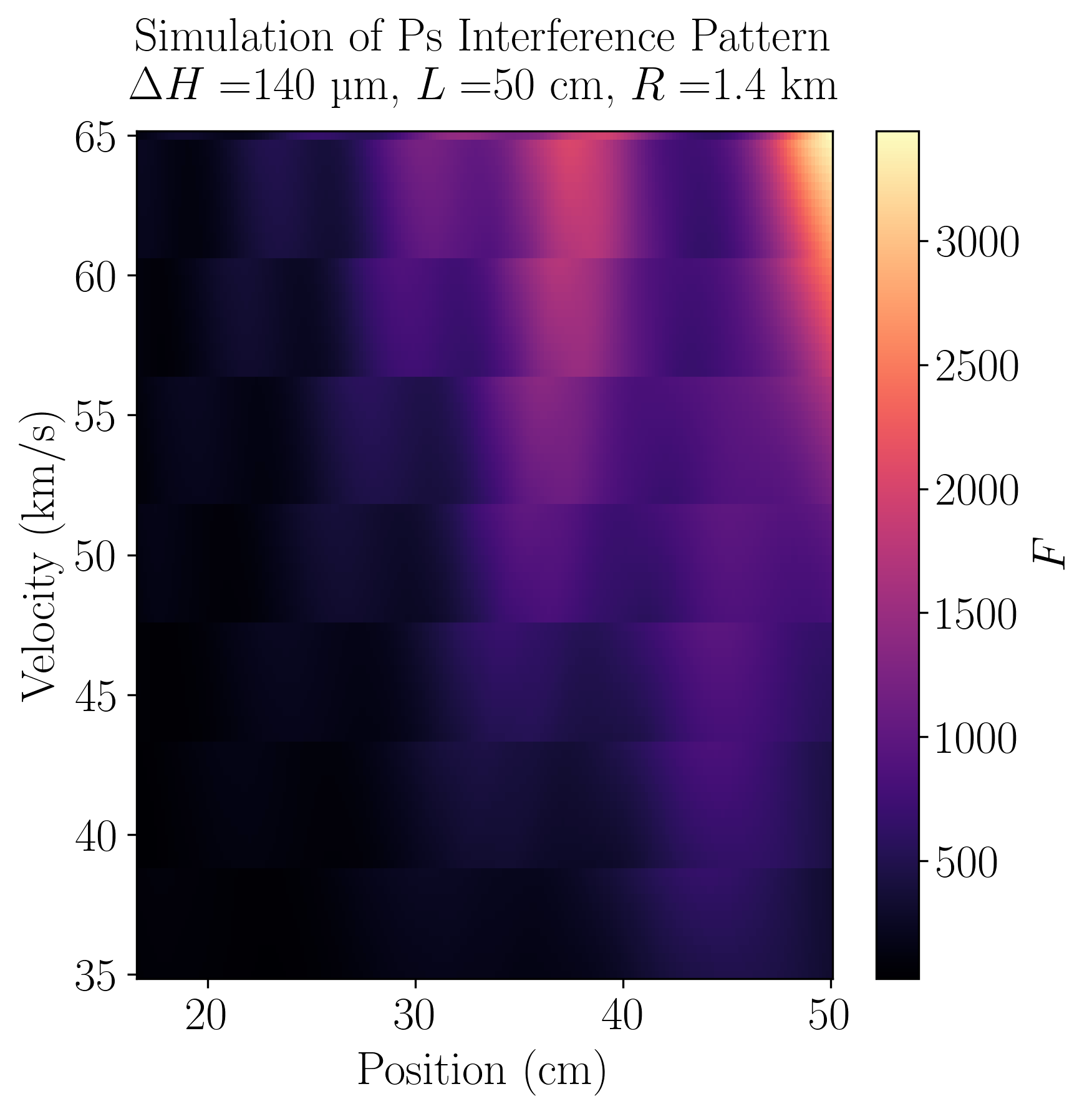}
    \caption{\label{fig:Ps_WG_Current} A simulation of the annihilation rate of \Positronium{} on  the surface of a mirror with the parameters in Eq. (\ref{eq:
    parameters for positronium}). Similar to the calculation of \Muonium{}, only the absorber/scatterer free region is shown. The absorber/scatterer has a length of $(L_A)_{WGS}^{Ps,n=33} = 16.6$~cm in this calculation. }
    \label{fig: Ps Current}
\end{figure}

The simulation results are presented in Fig. \ref{fig: Ps Current}.

\subsection{A scheme of a \Positronium{} WGS experiment}

Due to the same arguments as in the case of \Antihydrogen{}, the scheme of a \Positronium{} experiment is the same as the scheme of the \Antihydrogen{} experiment. In this case, a PET-like detector could be used to reconstruct the annihilation vertex of \Positronium{} on the mirror. 

\subsection{A possible improvement}

Although it seems feasible to measure the WGS of \Positronium{} and the QR can be studied, detection of the gravitational shift is much more difficult. The results shown in the previous sections are on the border between "very challenging" and "impossible". However, unlike \Muonium{}, whose lifetime cannot be changed, the lifetime of \Positronium{} can be, in principle, increased using excited states with $n>33$. This change may increase the relative magnitude of the gravitational shift (see Eq. \ref{eq: positronium gravitational shift}), which is proportional to $(\tau^{\Positronium})^{1.5}$ which is quite a strong factor. In this case, the gravitational shift could probably be measured as well.  

Another improvement is the use of slower \Positronium{}. This can be achieved by selecting slower \Positronium{} atoms from the broad velocity distribution, or simply sending the beam at a larger angle to the axis of the cylindrical mirror \footnote{Only the component of \Positronium{} velocity perpendicular to the cylinder axis matters.}. In this case, the mirror radius can decrease, and the production of optical elements of the experiment would become less challenging. Recent progress in laser cooling of positronium \cite{PhysRevLett.132.083402,Shu2024} holds great promise in this respect. 

A major difficulty that must be studied in detail is that the critical energy of all materials for \Positronium{} (and \Muonium{}) is "too" high. Moreover, \Positronium{} is reflected with a high probability not only due to QR but also in contact with the surface. Therefore, the probability of annihilation of \Positronium{} per quasi-classical bounce is very low. While \Positronium{} are reflected in the specular direction because of QR, the contact reflection might be off specular. 

\section{Conclusion}

We presented a general method for optimizing experiments that rely on various shifts of interference patterns formed by whispering-gallery and gravitational quantum states of various particles and atoms. Relatively small changes in the key parameters of the problem lead to changes in the WGS and GQS observables by many orders of magnitude, thus simplifying the optimization of experiments for solving specific problems. This approach has been applied to particles and atoms such as neutrons, hydrogen, antihydrogen, muonium, and positronium.  Our approach can be extended to other particles, other experimental arrangements, and combinations of them. We proposed specific parameters and experimental schemes to observe the whispering-gallery interference patterns and their shifts. In particular, we describe for the first time feasible experiments to observe and study whispering-gallery states of muonium and positronium. 

Note that each of the considered cases can be viewed as a "zero approximation" and a proof of the feasibility of the corresponding experiment. The sensitivity/accuracy of these experiments might be significantly increased by "tuning" the parameters of the problem, increasing the (spatial, temporal, energetic, etc.) resolutions, increasing the statistics and properties of the particle beams, etc. An interesting way to increase sensitivity is to use the revival phenomenon, which was analyzed in detail in relation to GQS and WGS in paper \cite{Robinett:2004}. The interference patterns already measured in neutron whispering-gallery experiments have a sufficiently complex structure to be observed and used.

The method developed has numerous applications. For example, the proposed neutron interferometer has high sensitivity and can be used to measure the gravitational constant, to improve the constraint on the electric charge of the neutron, and for other applications. The hydrogen/deuterium interferometer provides unprecedented sensitivity to the quantum reflection properties of hydrogen/deuterium atoms from surfaces. All presented methods can be extended to heavier atoms. The antihydrogen interferometer can be implemented with antihydrogen velocities of at least $\sim 10^5$~m/s. The identification and use of materials with an effective critical energy greater than that of silica would further increase the upper limit of the antihydrogen velocity range. Even a velocity of $\sim 10^6$~m/s is possible only with the use of flat liquid helium surfaces and extra magnetic fields. If the velocity is not significantly larger than $\sim 10^4$~m/s, measurements of a gravitational shift are experimentally feasible. Antihydrogen and positronium interferometers offer an advantage for studying quantum reflection in that antimatter particles are annihilated upon contact with a surface, thus preventing spurious effects. Positronium interferometers can in principle allow one to measure gravitational shifts, although extremely challenging absolute precision is needed. Neutron, atom, and antiatom interferometers, under reduced gravity, provide long-lived quantum states and exhibit significantly increased sensitivity and precision.

\begin{acknowledgments}
The authors thank all members of the GRASIAN, GRANIT, and GBAR collaborations for the fruitful and stimulating discussions. JP and SB are supported by the National Science Foundation under grant PHY-2412782. JP, KS, VN and EW are supported by AMADEUS under grant "Measurements of whispering gallery states (WGS) of neutrons and atoms".
\end{acknowledgments}

\nocite{*}

\bibliography{apssamp}

\appendix
\section{Stationary Phase Approximation}
\label{sec: Stationary Phase}

The stationary phase approximation is used to approximate Eq. (\ref{eq: paraxial beam1}) as Eq. (\ref{eq: far field wave packet}). The details of the approximation are worked out here. We can rewrite Eq. (\ref{eq: paraxial beam1}) as
\begin{equation}
\Psi(x,z) \approx \frac{1}{\sqrt{2\pi}}e^{ik_0z}\int_{-\infty}^{\infty}\tilde{\psi}(k)e^{i\phi(k)} dk \quad ,
\label{eq: paraxial beam2}
\end{equation}
where 
\begin{equation}
    \phi(k) = -\frac{k^2}{2k_0}z+kx \quad .
    \label{eq: phase}
\end{equation}
The phase rapidly oscillates for $k$ of the relevant scale describing the wave packet, so much of this integral cancels beyond the stationary point of $\phi(k)$. The stationary point of of Eq. (\ref{eq: phase}) occurs for $k=k_s$ with 
\begin{equation}
    k_s= k_0\frac{x}{z} \quad .
\label{eq: stationary k}
\end{equation}
Note also that
\begin{equation}
    \frac{\partial^2\phi}{\partial k^2}(k_s)=-\frac{z}{k_0} \quad .
    \label{eq: 2nd Derivative of phase}
\end{equation}
Using Eqs. (\ref{eq: phase} - \ref{eq: 2nd Derivative of phase}), the phase can be approximated about its stationary point as
\begin{equation}
\begin{split}
    \phi(k) \approx& \phi(k_s) + \frac{1}{2} \frac{\partial^2\phi}{\partial k^2}(k_s)(k-k_s)^2\\
    =& \frac{k_0}{z}\frac{x^2}{2}-\frac{z}{2 k_0}\left(k-k_0\frac{x}{z}\right)^2 \quad .
\end{split}
\end{equation}
If we approximate $\tilde{\psi}$ to zeroth order of $k_s$, we find the approximate position space wave function to be
\begin{equation}
\Psi(x,z)\approx\frac{1}{\sqrt{2\pi}}\tilde{\psi}
\left(k_0 \frac{x}{z}\right)e^{i\left(k_0z+\frac{k_0}{z}\frac{x^2}{2}\right)}\int_{-\infty}^{\infty} e^{-i \frac{z}{2k_0}\left(k-k_0\frac{x}{z}\right)^2} dk \quad . 
\end{equation}
By changing the coordinates from $k\rightarrow s$ where $s=\sqrt{\frac{z}{2k_0}}\left(k-k_0\frac{x}{z}\right)$ we find that the above integral becomes 
\begin{equation}
    \Psi(x,z)\approx \tilde{\psi}
    \left(k_0 \frac{x}{z}\right)e^{i\left(k_0z+\frac{k_0}{z}\frac{x^2}{2}\right)}\sqrt{\frac{k_0}{\pi z}}\int_{-\infty}^{\infty} e^{-is^2} ds \quad . 
\end{equation}
The above expression is the Fresnel integral and has the value
\begin{equation}
    \int_{-\infty}^{\infty} e^{-is^2} ds = \sqrt{\pi}e^{i\frac{\pi}{4}} \quad .
\end{equation}
and we find that
\begin{equation}
    \Psi(x,z)\approx\sqrt{\frac{k_0}{z}}\tilde{\psi}
    \left(k_0 \frac{x}{z}\right)e^{i\left(k_0z+\frac{k_0}{z}\frac{x^2}{2}+\frac{\pi}{4}\right)} \quad .
\end{equation}

\section{Sensitivity Estimate}
\label{sec: Sensitivity}
To estimate the sensitivity of the described experiments to an external force, we calculate the Cramér-Rao bound on the variance $\sigma^2_{\hat{a}}$ of the estimator $\hat{a}$ of the applied acceleration $a$. This method has been used to estimate the sensitivity of similar experiments in \cite{Crepin2019, Rousselle2022,Guyomard2025}. This estimation will assume that $\hat{a}$ is unbiased and efficient, an assumption that should be tested directly with Monte-Carlo simulations. This bound can be calculated as the inverse of Fisher information
\begin{equation}
\mathcal{I}=\mathbb{E}\left[\left( \frac{\partial \log{\mathcal{L}}}{\partial a} \right)^2 \right] \quad ,
\label{eq: Fisher Info1}
\end{equation}
where $\mathcal{L}$ is the likelihood function for an experiment that samples the probability distribution $P(X, V, a, \vec{\theta})$ $N$ times. $P$ corresponds to the simulated interference patterns, $X$ is the position, $V$ the velocity, and $\vec{\theta}$ are nuisance parameters. With the likelihood defined as 
\begin{equation}
\mathcal{L} = \prod_{i=1}^N P\left(X_i, V_i, a, \vec{\theta} \right) \quad ,
\label{eq: Likelihood}
\end{equation}
the Fisher information becomes
\begin{equation}
\begin{split}
    \mathcal{I} &= \int\cdots\int \sum_{i=1}^{N}{\left( \frac{\partial}{\partial a} \log{P(X_i, V_i, a, \vec{\theta})}\right)^2} \prod_{j=1}^NP(X_j, V_j, a, \vec{\theta}) dX_j dV_j\\
    &=  N \int\int \frac{\left(\frac{\partial}{\partial a} P(X, V, a, \vec{\theta})\right)^2}{P(X, V, a, \vec{\theta})} dX dV \quad ,  \\
\end{split}
\label{eq: Fisher Info2}
\end{equation}
where $X_i, V_i$ are the coordinates of the sampled event. The Cramér-Rao bound estimate on the error $\hat{a}$ is then
\begin{equation}
    \sigma_{\hat{a}} = \frac{1}{\sqrt{N}} \left( \int\int \frac{\left( \frac{\partial}{\partial a} P(X, V, a, \vec{\theta}) \right)^2}{P(X, V, a, \vec{\theta})} dX dV \right)^{-\frac{1}{2}} \quad .
\label{eq: Cramér Rao Bound}
\end{equation}

\end{document}